\documentclass[aps,pre,reprint,superscriptaddress]{revtex4-1}
\usepackage[dvipdfmx]{graphicx}
\usepackage{dcolumn}
\usepackage{bm}
\usepackage{color}
\begin{document}

\title{\centering{Identification of electrostatic two-stream instabilities associated with a laser-driven collisionless shock in a multicomponent plasma}}
\author{Youichi Sakawa}%
 \email{sakawa-y@ile.osaka-u.ac.jp}
\affiliation{%
Institute of Laser Engineering, Osaka University, Japan
}%

\author{Yutaka Ohira}
\affiliation
{Graduate School of Science, Tokyo University, Japan}%

\author{Rajesh Kumar}
\affiliation{Graduate School of Science, Osaka University, Japan}

\author{Alessio Morace}
\affiliation{Institute of Laser Engineering, Osaka University, Japan}

\author{
Leonard N. K. D$\ddot{\rm o}$hl}
\altaffiliation [Present address: ]
{Glen Eastman Energy b.v., Van Nelleweg 1, Expeditiegebouw, 3044 BC Rotterdam, The Netherlands}%
\affiliation
{York Plasma Institute, Department of Physics, University of York, UK}%

\author{Nigel Woolsey}
\affiliation
{York Plasma Institute, Department of Physics, University of York, UK}%

\date{\today}

%
\begin{abstract}
Electrostatic two-stream instabilities play essential roles in an electrostatic collisionless shock formation.
They are a key dissipation mechanism and result in ion heating and acceleration.
Since the number and energy of the shock-accelerated ions depend on the instabilities, precise identification of the active instabilities is important.
Two-dimensional particle-in-cell simulations in a multicomponent plasma reveal ion reflection and acceleration at the shock front, excitation of a longitudinally propagating electrostatic instability due to a non-oscillating component of the electrostatic field in the upstream region of the shock, and generation of up- and down-shifted velocity components within the expanding-ion components.
A linear analysis of the instabilities for a $\rm C_2H_3Cl$ plasma using the one-dimensional electrostatic plasma dispersion function, which includes electron and ion temperature effects, shows that the most unstable mode is the electrostatic ion-beam two-stream instability (IBTI), which is weakly dependent on the existence of electrons.
The IBTI is excited by velocity differences between the expanding protons and carbon-ion populations.
There is an electrostatic electron-ion two-stream instability with a much smaller growth rate associated with a population of protons reflecting at the shock.
The excitation of the fast-growing IBTI associated with laser-driven collisionless shock increases the brightness of a quasi-monoenergetic ion beam.
\end{abstract}

\maketitle
\section{Introduction}
In unmagnetized plasmas, a relative drift between two plasma populations results in the excitation of electrostatic two-stream instabilities.
In the case of cold ion beams, the electrostatic ion-beam two-stream instability (IBTI), which is a resonant instability driven by slow- and fast-ion beams, is excited \cite{Ohira2008}.
When a relative drift exists between electrons and ions, the Buneman instability is excited \cite{Buneman1963}.
By adding electrons with Maxwellian velocity-distribution function to the cold drifting-ions, the system is possibly unstable to the electrostatic ion-ion acoustic instability (ion-ion AI) and  the electrostatic electron-ion acoustic instability (electron-ion AI) \cite{Forslund1970,Karimabadi1991,Ohira2008}, in addition to IBTI and Buneman instability \cite{Ohira2008}.
The electron-ion AI and ion-ion AI are excited in the electron background.
Whereas the electron-ion AI is excited by the relative drift between electrons and ions, the ion-ion AI is excited when the relative drift between ion species is present.
Therefore, in a multi-species-ion plasma with relative drifts between ion species, both the  electron-ion AI and ion-ion AI can be excited in addition to IBTI.
It is well known that the growth rates of the electron-ion AI and ion-ion AI are much smaller than that of IBTI \cite{Ohira2008,Forslund1970}.

\citet{Akimoto1986} have conducted a linear analysis to study a broad-band electrostatic noise excited by an ion beam in the Earth's Magnetotail, and shown that the broad-band electrostatic noise can be explained by the presence of the ion-ion AI and electron-ion AI.
\citet{Wahlund1992} have revealed that the observation of enhanced ion-acoustic line spectra in the topside aurolar ionosphere results from the ion-ion AI or IBTI in a multicomponent (H$^+$, O$^+$, and NO$^+$) plasma.
The ion-ion AI and IBTI  have been observed in ion beam-plasma \cite{Gresillon1975,Ohnuma1976,TakaoFujita1977}
and laser-plasma \cite{Sarraf1983,Ross2013,Rinderknecht2018,Jiao2019} experiments.

Electrostatic two-stream instabilities play essential roles in collisionless shock formation as a dissipation mechanism, which results in ion acceleration and heating mechanism.
Two-dimensional (2D) particle-in-cell (PIC) simulations were conducted to investigate the ion-ion AI and IBTI in collisionless shocks \cite{Ohira2008,Karimabadi1991,Kato2010b,Sarri2011b,Zhang2018}.
\citet{Ohira2008} have investigated at the foot region of a collisionless shock with a very high Mach-number over 100, the fastest-growing mode is not the electron-ion AI but the highly-oblique ion-ion AI and IBTI excited by the shock-reflected ions.
\citet{Sarri2011b} have shown that the nearly-transverse ion-ion AI is excited by the reflected protons from the laser-driven electrostatic collisionless shock.
This work provides details of electrostatic two-stream instabilities associated with laser-driven collisionless shocks that occur in a multicomponent plasma.

Recently,  we reported a laser-driven electrostatic collisionless shock acceleration \cite{Denavit1992,Silva2004,Fiuza2012} of ions in multicomponent plasmas and excitation of electrostatic ion two-stream instabilities \cite{Kumar2019a}.
Hereafter, we refer to this work as Paper I.
The electrostatic collisionless shock acceleration has been demonstrated in the laboratory using a 10 $\mu$m wavelength CO$_2$ laser and a near-critical density gas target \cite{Haberberger2011}.
Several experiments on collisionless shock acceleration have been carried out in the last few years \cite{Tresca2015,Zhang2015,Zhang2017,Antici2017,Pak2018a,Ota2019}.
For medical applications, such as cancer therapies \cite{Bulanov2014a}, quasi-monoenergetic ions are preferred.  A laser-driven collisionless-shock-acceleration is a candidate ion source as the weak sheath field results in a quasi-monoenergetic ion beam \cite{Grismayer2006}.

In Paper I, Kumar {\it et al.} demonstrated using PIC simulations the possibility of producing high-flux and low energy-spread proton beams in a multicomponent C$_2$H$_3$Cl plasma.
The target consists of a tailored density profile with an exponentially decreasing density with 30 $\mu$m scale-length on the rear side. This results in a uniform electrostatic sheath field $E_{\rm TNSA}$ ahead, upstream, of the shock. Expansion of ions under $E_{\rm TNSA}$ results in relative drifts between slower-moving C ions with  lower average charge-to-mass ratio $\langle Z \rangle / \langle A \rangle = 0.5$ and faster-moving protons with higher $\langle Z \rangle / \langle A \rangle = 1$.
It was shown that the development of the longitudinal electrostatic ion two-stream instabilities play important roles in multicomponent plasmas and the associated ion acceleration process.
By using the cold-ion approximation and ignoring the electrons, it was shown that two electrostatic ion two-stream instabilities or two IBTIs can be excited: One is the heavy-ion electrostatic ion two-stream instability, which is excited between the expanding proton and C-ion populations.
This instability occurs in a multicomponent plasma as ion components have different $\langle Z \rangle / \langle A \rangle$ ratios, such as in a CCl$_2$ plasma with fully ionized C$^{6+}$ ($\langle Z \rangle / \langle A \rangle = 0.5$) and Cl$^{10+}$ ($\langle Z \rangle / \langle A \rangle = 0.28$) ions as shown in Fig.~7(b) of Paper I.
The other is reflected-proton electrostatic ion two-stream instability, which is excited between the reflected and expanding proton populations associated with the shock.
In this analysis, the growth rate and the wavenumber of the most unstable modes of the instabilities are derived from an analytical model with two cold-ion populations without a treatment of the electron population.

In Ref.~\cite{Kumar2021}, we observed two electrostatic collisionless shocks at two distinct longitudinal positions when driven with laser at normalized laser vector potential $a_0>10$.
Moreover, these shocks, associated with protons and carbon ions accelerate ions to different velocities in an expanding upstream with higher flux than in a single-component hydrogen or carbon plasma.
A broadening upwards of the C$^{6+}$-ion velocity distribution, which is important to increase the number of the accelerated C$^{6+}$ ions, is predicted to result from the heavy-ion electrostatic ion two-stream instability \cite{Kumar2021}. 

In this paper, we report on the identification of electrostatic two-stream instabilities associated with laser driven electrostatic collisionless shocks in a multicomponent C$_2$H$_3$Cl plasma investigated using 2D PIC simulations.
These PIC simulations use the normalized laser vector potential $a_0=3.35$, as discussed in Paper I, to investigate the electrostatic ion two-stream instability.
At this laser intensity, only a proton shock is excited.
A linear analysis of the instabilities for a $\rm C_2H_3Cl$ plasma, with electrons, protons, C and Cl ions, is carried out using the one-dimensional electrostatic plasma dispersion function for unmagnetized collisionless plasmas including ion temperature effect to identify the electrostatic ion two-stream instability.
We use plasma parameters, such as temperatures, densities, and drift velocities for all the species of particles, obtained from the PIC simulations. To identify the instabilities, we start the linear analysis from the case of cold ions and ignoring the role of electrons, in which IBTI can be excited, and artificially removing some ion species.
This is extended to cold ions and hot electrons, in which the electron-ion AI and ion-ion AI are excited.
Finally, finite-temperature ions are included to understand the influence of the ion Landau damping.

The paper is structured as follows: In Sec.~II, we discuss the EPOCH \cite{Arber2015} 2D PIC calculations in a multicomponent $\rm C_2H_3Cl$ plasma. We describe the temporal evolution of the proton phase-space and the broadening of upstream expanding-proton distribution.
Sec.~III outlines a linear instability analysis using data from the numerical simulations.
This section is divided into four parts and examines the instabilities for cold ions without electrons (Part.~III-A), cold ions with hot electrons (Part.~III-B), warm ions with hot electrons (Part.~III-C), and in Part.~III-D we identify the observed instabilities.
The results of the numerical simulations and the linear analysis are discussed in Sec.~IV and summarized in Sec.~V.

\section {Particle-in-cell Simulation in a multicomponent plasma}

The EPOCH calculations are conducted with the same parameters as described in Paper I.
The simulation box is 300 $\mu$m $\times$ 6 $\mu$m in size and composed of 9000 $\times$ 180 cells along the $x$- and $y$-axis respectively, with 30 particles per cell.
The skin depth is resolved by 2.7 cells, and the electron-proton mass ratio of 1836 is used.
The boundary conditions are open in the $x$-direction and periodic in the $y$-direction for both fields and particles. The laser pulse is modeled as the electromagnetic plane wave with linear p-polarization along the $y$-axis and propagates in the $x$-direction. The normally incident laser pulse with infinite spot size has a Gaussian temporal profile with 1.5 ps full-width at half-maximum. The peak intensity is 1.4$\times$10$^{19}$ W/cm$^2$ (a$_0$ = 3.35). The laser pulse interacts with a fully ionized plasma density at $x$ = 40 $\mu$m. 
We use a $\rm C_2H_3Cl$ plasma with a longitudinal ($x$-direction) density profile consisting of an exponentially increasing $5 ~\mathrm{\mu m}$ scale-length laser-irradiated front region, $5 ~\mathrm{\mu m}$ uniform central region, and an exponentially decreasing profile with $30 ~\mathrm{\mu m}$ scale-length rear region as the back of the target. To avoid boundary effects, the simulations use 40 $\mu$m ($x = 0 - 40$ $\mu$m) and 100 $\mu$m ($x = 200 - 300$ $\mu$m) vacuum regions at the front and rear of the target, respectively.  Details of the simulations including the target density profiles at $a_0 = 3.35$ are given in \citet{Kumar2019a,Kumar2021}. 
The maximum electron density is fixed at the relativistic critical density a$_0n_{cr}$, where $n_{cr}$ = 1.12$\times$10$^{21}$ cm$^{-3}$ is the critical plasma density for the 1.053 $\mu$m wavelength laser used in these simulations. 
The charge states $Z$ of protons, C ions, and Cl ions are $1$, $6$, and $15$, respectively.
Cl (atomic number 17) is ionized to the He-like ion state, Cl$^{15+}$. The three ions have average charge-to-mass ratios $\langle Z \rangle / \langle A \rangle$ of 1, 0.5 and 0.42 respectively.
The corresponding ion density for each material is calculated from the quasi-neutral plasma condition.
Initial temperatures of particles are 500 eV for all species.
%
  \begin{figure}[b]
  \includegraphics[width=\linewidth]{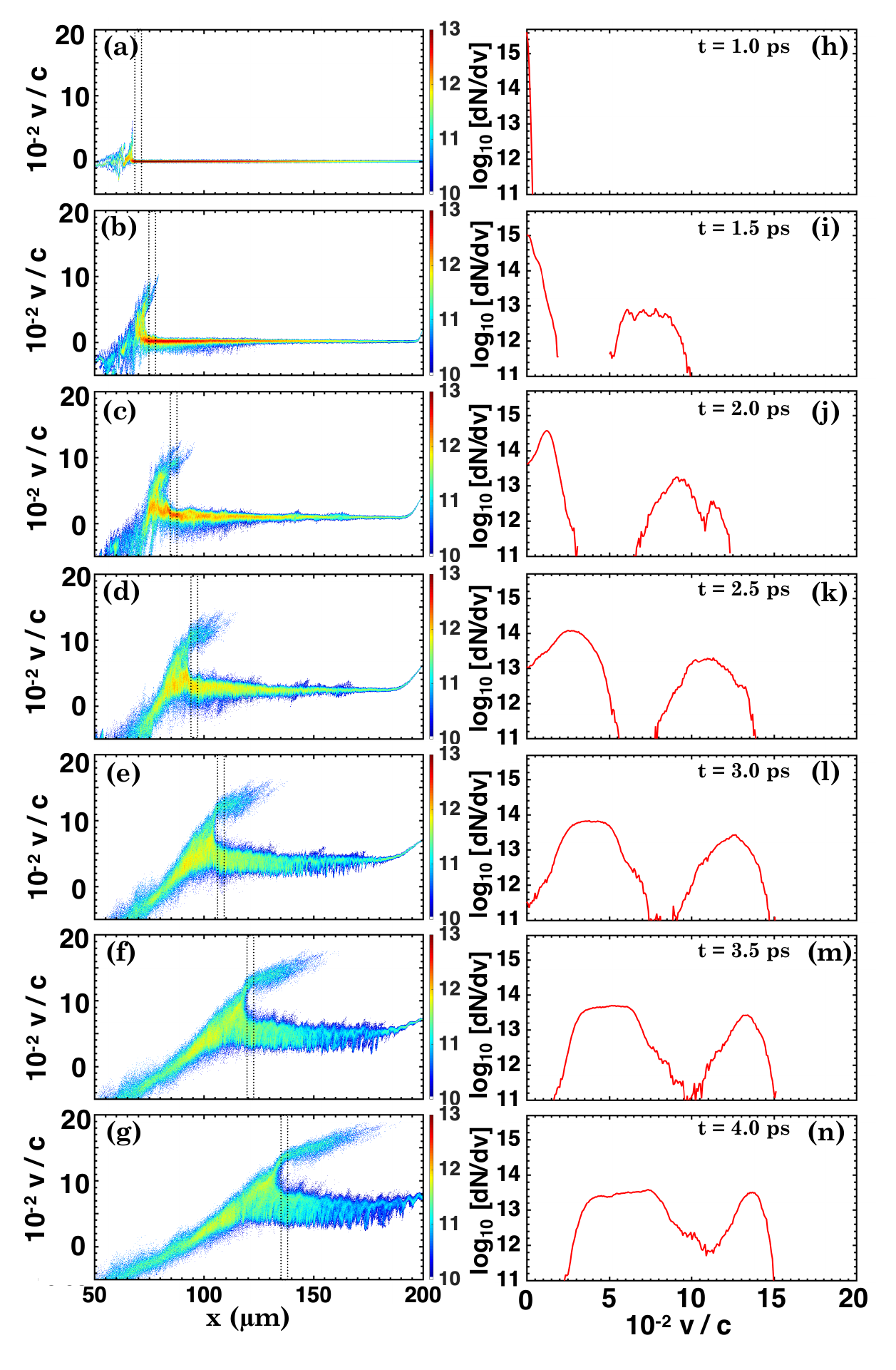}
  \caption{The temporal evolution of (a)-(g) proton phase-space and (h)-(n) their corresponding velocity spectrum taken at $\Delta x$ = 3 $\mu$m in the upstream region (shown by vertical dotted lines in phase-space) in a $\rm C_2H_3Cl$ plasma. The color scale shows the number of ions in a log scale.}
   \label{fig:11}
\end{figure}

To accelerate the ions via the collisionless shock acceleration mechanism, the potential energy at the shock front must be larger than the kinetic energy of the upstream expanding ions in the shock rest frame. In other words, the electrostatic potential at the shock front must satisfy the following conditions \cite{Tidman}, $Ze\phi \geq \frac{1}{2} Am_p(V_{sh}^i-v_0^i)^2$.
Here, $\phi$,  $V_{sh}^i$, and $v_0^i$ are the electrostatic potential, the shock velocity, and the particle velocity, superscript $i$ represents the different ion species. The lower threshold ($v_L^i$) in $v_0^i$ for ion reflection via the collisionless shock acceleration mechanism is $v_L^i = V_{sh}^i - \sqrt{2(Z_i /A_i) e\phi / m_p}$. 
Namely, the reflection condition is given by $v_L^i \leq v_0^i \leq V_{sh}^i$ \cite{Kumar2019a,Kumar2021}.
This equation represents the lower $v_L^i$ and upper $V_{sh}^i$ bounds in $v_{0}^i$ for ion reflection. Therefore, all ions with a velocity component in between $v_L^i$ and $V_{sh}^i$ are reflected by the shock potential to a velocity  $V_{sh}^i +  \sqrt{2(Z_i /A_i) e \phi / m_p}$ = 2$V_{sh}^i-v_L^i$. 
For protons $Z_i$ = $A_i$ = 1 and neglecting the superscript $i$, $v_L = V_{sh} - \sqrt{2e\phi /m_p}$. 

Figure \ref{fig:11} represents the temporal evolution of the proton phase-space and the corresponding velocity spectrum taken at $\Delta x$ = 3 $\mu$m shown by the vertical lines on the phase-space. A significantly large number of protons satisfy the reflection condition at all times. Therefore, a large fraction of protons is accelerated via the collisionless shock acceleration mechanism.

\begin{figure}
  \includegraphics[width=\linewidth]{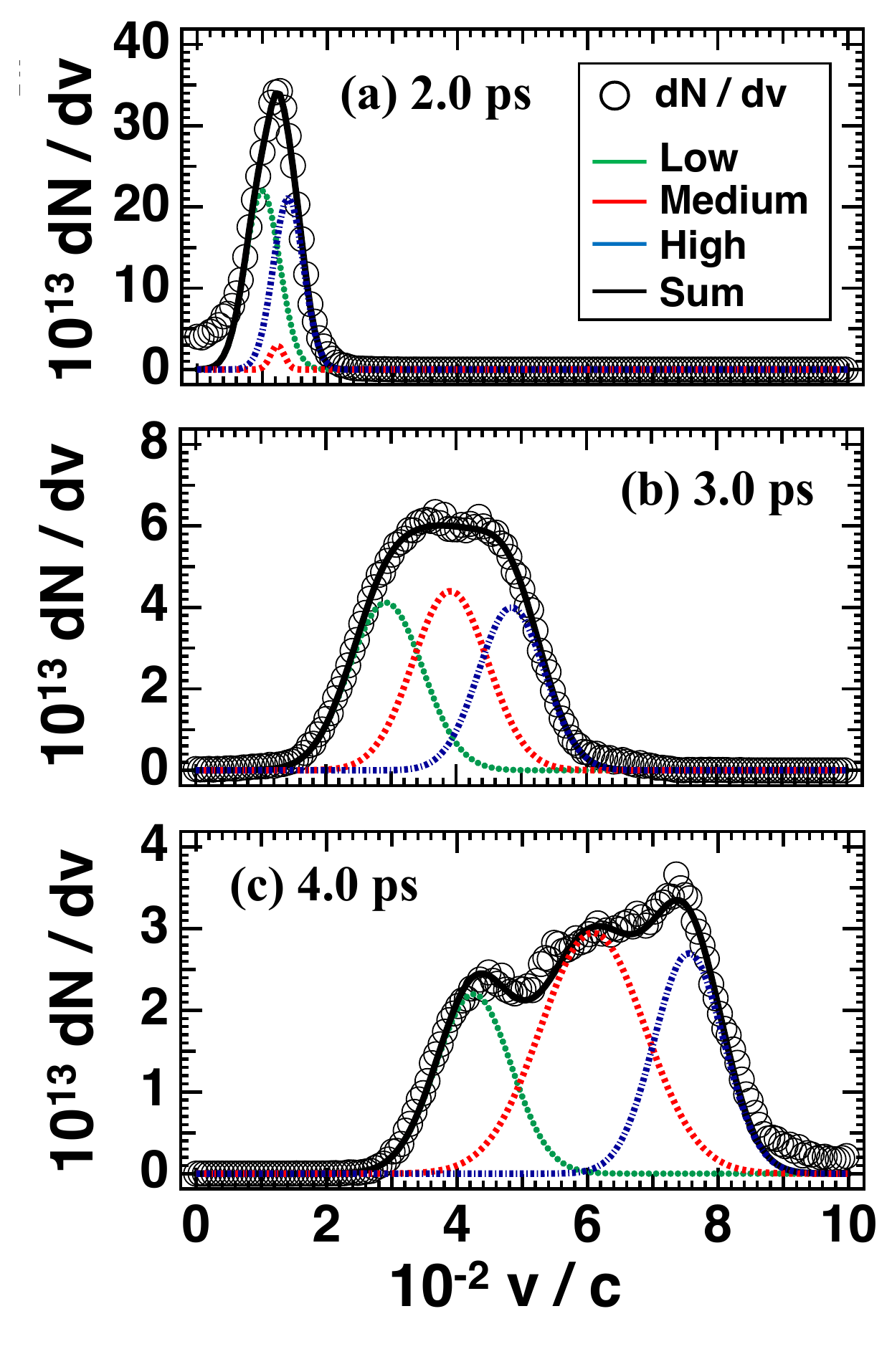}
  \caption{The velocity spectrum (open circles) of the upstream expanding protons taken at $\Delta x$ = 3 $\mu$m of the upstream region in a $\rm C_2H_3Cl$ plasma at (a) \textit{t} = 2.0, (b)  3.0, and (c)  4.0 ps, which is replotted from Fig.~\ref{fig:11}. Note that the $y$ axis is in the linear scale. A sum (black line) of the three 1D shifted-Maxwellian distributions [low (green line), medium (red line), and high (blue line) velocity components] are used to fit the upstream expanding protons.}
   \label{fig:8}
\end{figure}

Figure \ref{fig:8} shows the temporal evolution of the upstream expanding-proton velocity distribution taken at $\Delta x$ = 3 $\mu$m in the upstream region of a $\rm C_2H_3Cl$ plasma, which is replotted from Fig.~\ref{fig:11}. The broadening of the upstream expanding-protons in a multicomponent plasma is caused by the two-stream instability as explained previously and in Paper I. The width of the velocity distribution in the $x$-direction increases with time as shown in Figs.~\ref{fig:8}(a)-\ref{fig:8}(c).
The shape of the velocity spectrum can be described by using three 1D shifted-Maxwellian distributions for all times, and examples are shown in Fig.~\ref{fig:8} at \textit{t} = 2.0, 3.0, and 4.0 ps.
%
%
%
\begin{figure}
\includegraphics[width=\linewidth]{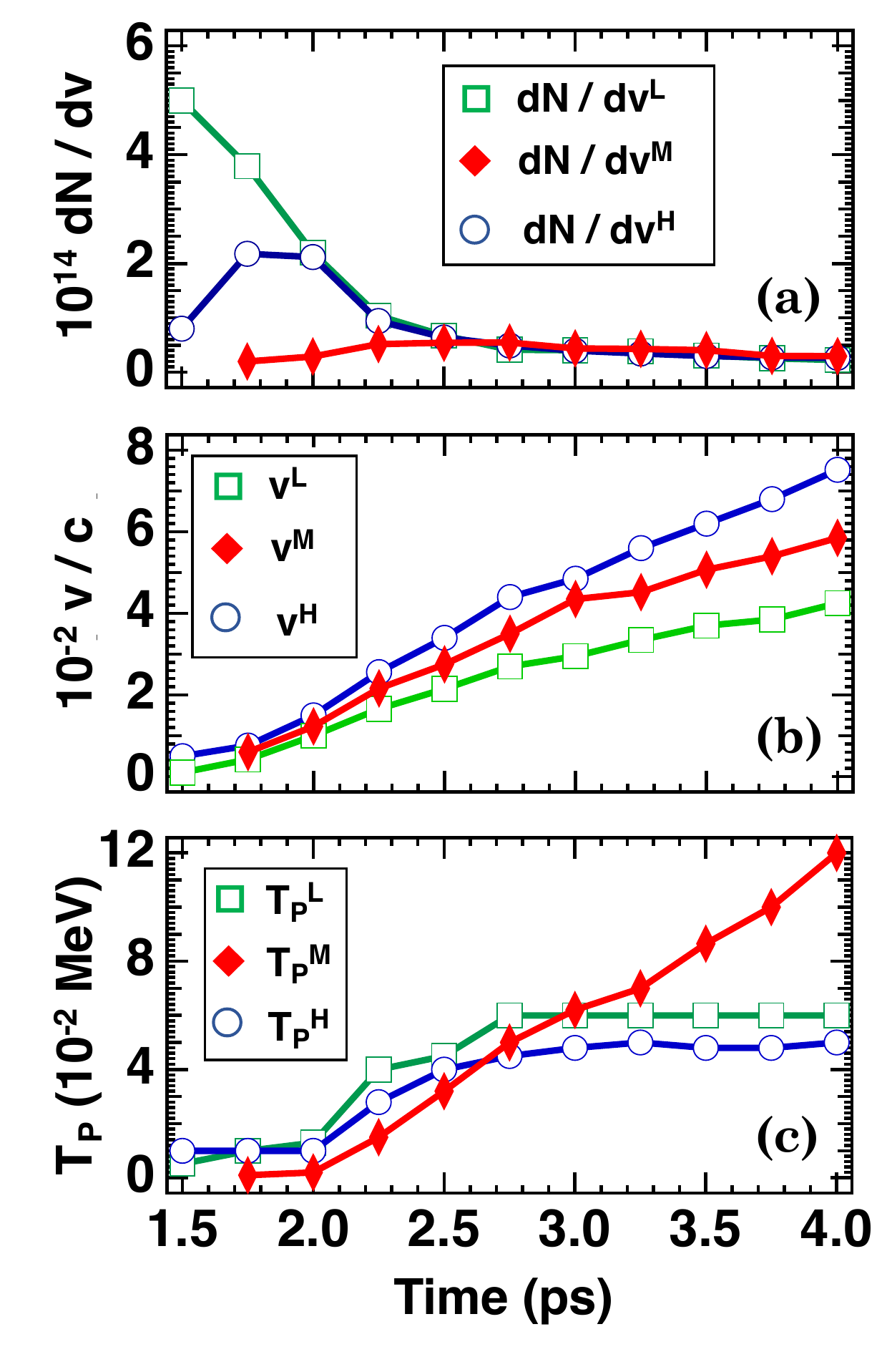}
\caption{Temporal evolution of the upstream expanding protons taken at $\Delta x$ = 3 $\mu$m of the upstream region in a $\rm C_2H_3Cl$ plasma. (a) Peak number density $dN/dv$, (b) velocities $v/c$ at the peak number density, and (c) proton temperature $T_{\rm P}$ of the low (L), medium (M), and high (H) velocity components of the fitted 1D shifted-Maxwellian distributions.}
\label{fig:9}
\end{figure}

Temporal evolution of the peak number density $dN/dv$, velocity $v$ at the peak number density, and the proton temperature $T_{\rm P}$ of the three fitted-Maxwellian distributions of the upstream expanding protons are represented in Figs.~\ref{fig:9}(a), \ref{fig:9}(b), and \ref{fig:9}(c), respectively.
In Fig.~\ref{fig:9}(a), the $dN/dv^{\rm L}$, $dN/dv^{\rm M}$, and $dN/dv^{\rm H}$ denote number densities at the peaks of the fitted-Maxwellian distributions for the low-,  medium-, and high-velocity components of the distributions in the upstream region.
It is clear from Fig.~\ref{fig:9}, at \textit{t} = 1.50 ps of the laser peak, a large number of protons are in the low-velocity component, a small number of protons are in the high-velocity component, and their temperatures are nearly the same ($T_{\rm P}^{\rm H} \simeq T_{\rm P}^{\rm L}  \simeq 0.01$ MeV).
After the laser peak has passed, protons in the upstream region are accelerated by a uniform sheath electric field $E_{\rm TNSA}$, and the peak velocities of three Maxwellian components keep increasing with time as shown in Fig.~\ref{fig:9}(b).
At $t$ = 1.75 ps, the third Maxwellian distribution component with a very low number density and low temperature starts to appear.
Temperatures of three velocity components start increasing at $t=2.0$ ps, and $T_{\rm P}^{\rm L}$ and $T_{\rm P}^{\rm H}$ remain the same with 0.06 and 0.048 MeV, respectively, after $t$ = 2.75 ps, while $T_{\rm P}^{\rm M}$ keeps increasing as shown in Fig.~\ref{fig:9}(c).
These increments in $T_{\rm P}^{\rm L}$ and $T_{\rm P}^{\rm H}$ show a similar trend as that in the electron temperature $T_e$, this is illustrated in Fig.~\ref{fig:4}(c) of the Appendix, with a delay of $\simeq 1$ ps, that is $T_e$ and $T_{\rm P}$ start increasing at 1 and 2 ps, respectively.
By $t$ = 2.75 ps peak values of $dN/dv^{\rm L}$, $dN/dv^{\rm M}$, and $dN/dv^{\rm H}$ become nearly equal, and they are  nearly constant later as shown in Fig.~\ref{fig:9}(a). 

Paper I reports on the formation of a  high-energy tail in expanding C$^{6+}$ ions in addition to the heating of expanding protons.
This heating of expanding protons and C$^{6+}$ ions result from the excitation of longitudinally propagating electrostatic two-stream instabilities.
\section{Linear analysis of instabilities}
In Paper I, using an approximated dispersion relation we have shown that the excitation of electrostatic instability leads to the broadening of upstream expanding-proton distribution in the multicomponent plasma.
The PIC simulations in Paper I show that the propagation direction of the electrostatic instability is longitudinal to the flow direction ($x$-direction).
However, the previous analysis did not take into account thermal effects.
In this section, to clarify the physical picture of the instability, we carry out a linear analysis of the longitudinal electrostatic instability for a $\rm C_2H_3Cl$ plasma using the one-dimensional electrostatic plasma dispersion function \cite{Ohira2008}.

For unmagnetized collisionless plasmas, the electrostatic dispersion relation is expressed as \cite{Ohira2008}
\begin{equation}
1+  \sum_{s = e + i} \frac{2 \omega_{{\rm p}s}^2} {k^2 v_{{\rm th}s}^2} [1 + \xi_s Z(\xi_s)]=0, 
\label{eq:disp}
\end{equation}
\begin{equation}
Z(\xi_s) = \frac{1}{\sqrt {\pi}} \int_{- \infty}^{+ \infty} \frac{e^{-z^2}}{z -\xi_s } dz,
\label{eq:Z}
\end{equation}
\begin{equation}
\xi_s =  \frac{\omega - k v_{s}}{k v_{{\rm th}s}}, 
\label{eq:xi}
\end{equation}
where $s$ is electron ($e$) and ion species ($i$); $\omega$ is the frequency;  $k$ is the wavenumber in the $x$-direction; $\omega_{{\rm p}s}$, $v_s$, and $v_{{\rm th}s}= \sqrt {2 T_s/ m_s}$ are the plasma frequency, drift velocity, and thermal velocity of particle species $s$; and $Z(\xi_s)$ is the plasma dispersion function \cite{Fried1961book}.
In the calculation, electrons ($s$ = $e$), expanding $\rm C^{6+}$ ions ($s$ = C), expanding $\rm Cl^{15+}$ ions ($s$ = Cl), expanding protons ($s$ = P-exp), and reflected protons ($s$ = P-ref) are included.
The electrostatic dispersion relation [Eq.~(\ref{eq:disp})] can be numerically solved.

When $T_i = 0$,  $v_e =0$, and $\xi_e \gg 1$, Eq.~(\ref{eq:disp}) is reduced to 
\begin{equation}
1 -\frac{\omega_{{\rm p}e}^2} {\omega^2 - k^2 v_{{\rm th}e}^2} = \sum_{i} \frac{\omega_{{\rm p}i}^2} {(\omega - k v_{\rm i})^2}.
\label{eq:cold_ion_w}
\end{equation}
On the other hand, when $T_i = 0$,  $v_e =0$, and $\xi_e \ll 1$, Eq.~(\ref{eq:disp}) is reduced to
\begin{equation}
1 + \frac{1}{k^2 \lambda_{De}^2} = \sum_{i} \frac{\omega_{{\rm p}i}^2} {(\omega - k v_{i})^2},
\label{eq:cold_ion}
\end{equation}
where, $\lambda_{De}$ is the electron Debye length. Equation (\ref{eq:cold_ion}) is rewritten as,
\begin{equation}
1 = \sum_{i} \frac{1}{1+ k^2 \lambda_{De}^2}\frac{ (n_i / n_e)k^2 (T_e / m_i) } {(\omega - k v_{i})^2}.
\label{eq:cold_ion2}
\end{equation}
When $1/ (k^2 \lambda_{De}^2)=0$, that is equivalent to no electron effect or $n_e = 0$, Eq.~(\ref{eq:cold_ion}) is reduced to 
\begin{equation}
1 = \sum_{i} \frac{\omega_{{\rm p}i}^2} {(\omega - k v_{i})^2}.
\label{eq:no_e}
\end{equation}

Table \ref{table:1} summarizes all the plasma parameters used in the analysis.
For electrons, relativistic plasma frequency of $\omega_{{\rm p}e}$ = $ \sqrt {n_e e^2 / \epsilon_0 \gamma m_e}$,  $v_{\rm {th}e}= \sqrt {2 T_e/ \gamma m_e}$,  and $\gamma =   3 T_e /  m_e c^2 =12$ are used, where $\epsilon_0$ is the vacuum permittivity and $c$ is the speed of light.
\begin{table}[h]
\caption{\label{tab:table1} Densities, plasma frequencies, temperatures, drift velocities, and relative drift velocities to Cl$^{15+}$ of particle species $s$ used in the linear analysis of the electrostatic two-stream instability for a $\rm C_2H_3Cl$ plasma.
Electrons ($s$ = $e$), expanding $\rm C^{6+}$ ions ($s$ = C), expanding $\rm Cl^{15+}$ ions ($s$ = Cl), expanding protons ($s$ = P-exp), and reflected protons ($s$ = P-ref) are included.
These parameters are derived from the 2D PIC simulations at $t = 4.0$ ps shown in the previous section. 
}
\label{table:1}
\centering
\begin{ruledtabular}
\begin{tabular}{l l l }
\multicolumn{1}{c}{\textbf{ }} & \textbf{Definition} &    \\
\hline
Density : $n_s$\\
\hspace {2mm}Electron & $n_e$ &  $(7.00 \pm 0.15) \times 10^{20}$ cm$^{-3}$  \\
\hspace {2mm}Expanding proton &$n_{\rm P-exp}$ &  $(9.40 \pm 0.22) \times 10^{19}$ cm$^{-3}$  \\
\hspace {2mm}Reflected proton &$n_{\rm P-ref}$ & $(3.10 \pm 0.08) \times 10^{19}$ cm$^{-3}$ \\
\hspace {2mm}Expanding C$^{6+}$ &  $n_{\rm C}$ & $(4.50 \pm 0.03) \times 10^{19}$ cm$^{-3}$   \\
\hspace {2mm}Expanding Cl$^{15+}$   & $n_{\rm Cl}$ & $(2.10 \pm 0.07) \times 10^{19}$ cm$^{-3}$ \\
Plasma frequency : $\omega_{{\rm p}s}$\\
\hspace {2mm}Electron  & $\omega_{{\rm p}e}$ & $4.3 \times 10^{14}$ s$^{-1}$  \\
\hspace {2mm}Expanding proton & $\omega_{\rm pP-exp}$ & $1.3 \times 10^{13}$ s$^{-1}$  \\
\hspace {2mm}Reflected proton & $\omega_{\rm pP-ref}$ & $7.4 \times 10^{12}$ s$^{-1}$  \\
\hspace {2mm}Expanding  C$^{6+}$  & $\omega_{\rm pC}$ &$1.5 \times 10^{13}$ s$^{-1}$  \\
\hspace {2mm}Expanding Cl$^{15+}$  & $\omega_{\rm pCl}$ &  $1.5 \times 10^{13}$ s$^{-1}$  \\
Temperature : $T_s$   \\
\hspace {2mm}Electron & $T_e$ & 2.0 MeV  \\
\hspace {2mm}Ion & $T_i$ & 2$\times$$10^{-5}$   \\
\hspace {2mm} &  & \hspace {2mm} - 0.4 MeV  \\
Drift velocity : $v_s$\\
\hspace {2mm}Electron & $v_e$ &   $0.042c$   \\
\hspace {2mm}Expanding proton &$v_{\rm P-exp}$ &  0.075$c$  \\
\hspace {2mm}Reflected  proton &$v_{\rm P-ref}$ &  0.139$c$  \\
\hspace {2mm}Expanding C$^{6+}$  &$v_{\rm C}$ &   0.033$c$ \\
\hspace {2mm}Expanding Cl$^{15+}$  & $v_{\rm Cl}$ & 0.03$c$   \\
Relative drift velocity \\ \hspace {3mm} to Cl$^{15+}$ : $v_{{\rm d}s}$ \\
\hspace {2mm}Expanding proton &$v_{\rm dP-exp} $ &  0.42$v_0$=0.045$c$  \\
\hspace {2mm}Reflected  proton &$v_{\rm dP-ref}$ & $v_0$=0.109$c$  \\
\hspace {2mm}Expanding C$^{6+}$  &$v_{\rm dC}$ &   0.028$v_0$=0.003$c$ \\
\hline
\end{tabular}
\end{ruledtabular}
\end{table}

In the following, we show the results of linear analysis of the electrostatic instability excited in a multicomponent $\rm C_2H_3Cl$ plasma by solving the dispersion relation [Eq.~(\ref{eq:disp})].
We represent 3 cases from the simplest to something more realistic: A. cold ions without electrons, B. cold ions with hot electrons, and C. finite-temperature ions.
In case A, the dispersion relation is approximated by Eq.~(\ref{eq:no_e}), which corresponds to the cold ion-beam interaction, and the excitation of the electrostatic ion-beam two-stream instability (IBTI) is shown.
In case B, the dispersion relation is approximated by Eqs.~(\ref{eq:cold_ion}) and (\ref{eq:cold_ion2}), and the excitation of  the electrostatic electron-ion acoustic instability (electron-ion AI) and the electrostatic ion-ion acoustic instability (ion-ion AI) is displayed.
In case C, the full dispersion relation [Eq.~(\ref{eq:disp})] is used, and the ion Landau damping effect and the ion-temperature dependence of the instability threshold are displayed.
Finally, we identify the instabilities observed in the PIC simulation.
The dispersion relations are expressed in a rest frame where the drift velocity of expanding Cl$^{15+}$ ions ($v_{\rm Cl}$) is zero.

%
\subsection{Cold ($T_i = 2 \times 10^{-5}$ MeV) ions and without electrons:  excitation of the electrostatic ion-beam two-stream instability (IBTI)}

%
 \begin{figure}
  \includegraphics[width=0.49\textwidth]{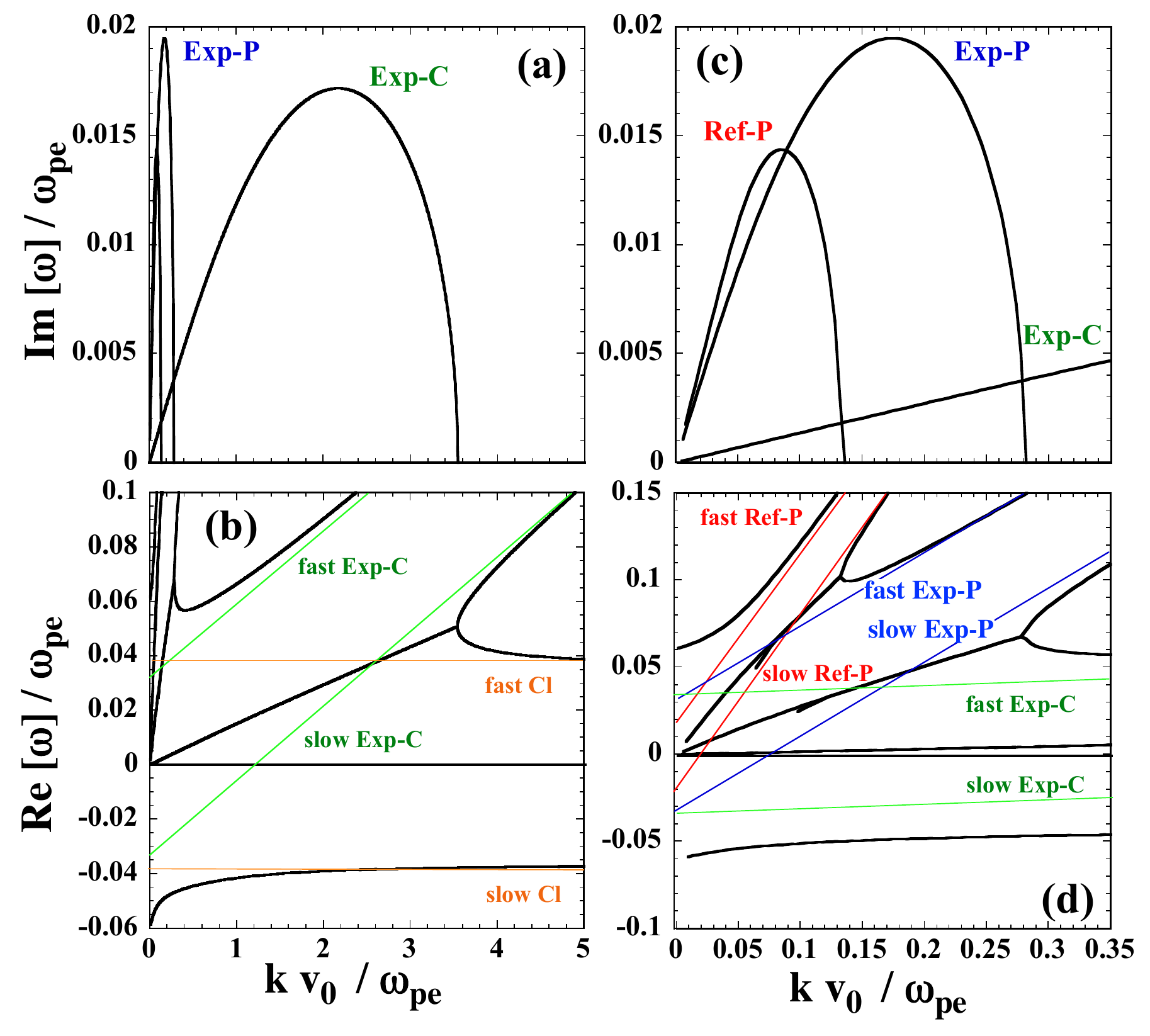}
  \caption{Results of the linear analysis of the electrostatic two-stream instability for a $\rm C_2H_3Cl$ plasma when $n_e = 0$ and  $T_i = 2 \times 10^{-5}$ MeV. 
(a) , (c) The imaginary part (Im [$\omega]/ \omega_{{\rm p}e}$) and (b), (d) real part  (Re [$\omega]/ \omega_{{\rm p}e}$) of normalized frequencies versus the normalized wavenumber in the $x$-direction ($k v_0 / \omega_{{\rm p}e}$, where $v_0 = v_{\rm P-ref} - v_{\rm Cl}$).  (a), (b) [(c), (d)] are plotted for $k v_0 / \omega_{{\rm p}e} < 5$ [$k v_0 / \omega_{{\rm p}e} <0.35$]. The thin red, blue, green, and orange lines in (b) and (d) are the fast and slow Ref-P, Exp-P, Exp-C, and Cl modes, respectively.}
   \label{fig:EITI_noNe_noTi}
\end{figure}
%
%
 \begin{figure}
  \includegraphics[width=0.43\textwidth]{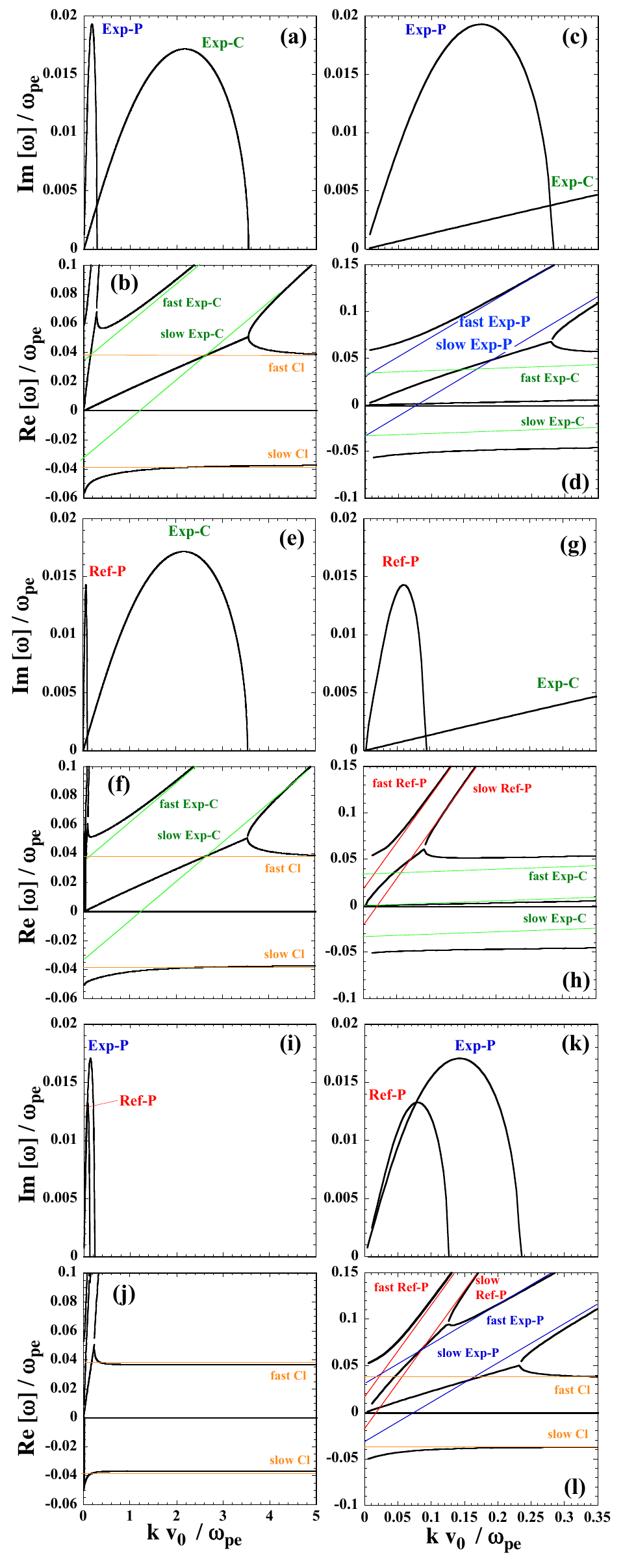}
  \caption{Ion species effect of the electrostatic two-stream instability for a $\rm C_2H_3Cl$ plasma when $n_e = 0$ and $T_i = 2 \times 10^{-5}$ MeV. 
(a), (c), (e), (g), (i), (k) The imaginary part (Im [$\omega]/ \omega_{{\rm p}e}$) and (b), (d), (f), (h), (j), (l)  real part (Re [$\omega]/ \omega_{{\rm p}e}$) of normalized frequencies versus the normalized wavenumber in the $x$-direction ($k v_0 / \omega_{{\rm p}e}$). The thin red, blue, green, and orange lines in (b) and (d) are the fast and slow Ref-P, Exp-P, Exp-C, and Cl modes, respectively. 
(a), (b), (e), (f), (i), (jl)[(c), (d), (g), (h), (k), (l)] are plotted for $k v_0 / \omega_{{\rm p}e} < 5$ [$k v_0 / \omega_{{\rm p}e} <0.35$].
(a), (b), (c), (d)  $n_{\rm P-ref} = 0$;  (e), (f), (g), (h) $n_{\rm P-exp} = 0$; and  (i), (j), (k), (l) $ n_{\rm C} = 0$.}
   \label{fig:EITI_iondep}
\end{figure}

First, we show the results of linear analysis for cold ions and no electron effects.
By applying this approximation [$1/ (k^2 \lambda_{De}^2) =0$ or equivalent to $n_e = 0$], the dispersion relation is reduced to Eq.~(\ref{eq:no_e}). The resultant instability is the ion-beam two-stream instability (IBTI) \cite{Ohira2008}.
This is a resonance instability driven by slow- and fast-ion beams with relative drifts between ion species.
The electron effect is negligible because of the large electron Debye length. 

Figures \ref{fig:EITI_noNe_noTi}(a) and \ref{fig:EITI_noNe_noTi}(b) show, respectively, normalized imaginary part or growth rate (Im [$\omega]/ \omega_{{\rm p}e}$) and real part (Re [$\omega]/ \omega_{{\rm p}e}$) of the instability frequency versus normalized wavenumber in the $x$-direction ($k v_0 / \omega_{{\rm p}e}$) when $T_i = 2 \times 10^{-5}$ MeV and  $n_e =0$.
Here, $\omega_{{\rm p}e}$ is the electron plasma frequency, and $v_0$ is the drift velocity $v_{\rm P-ref}$ of reflected protons in a rest frame where the drift velocity $v_{\rm Cl}$ of expanding Cl$^{15+}$ ions is zero ($v_0 = v_{\rm dP-ref} = v_{\rm P-ref} - v_{\rm Cl} = 0.109c$). 
When $T_i = 2 \times 10^{-5}$ MeV, the ion temperature effect is negligible, and the dispersion relation is reduced to Eq.~(\ref{eq:no_e}).
Figures \ref{fig:EITI_noNe_noTi}(c) and \ref{fig:EITI_noNe_noTi}(d) are replotted from Figs.~\ref{fig:EITI_noNe_noTi}(a) and \ref{fig:EITI_noNe_noTi}(b), respectively, for $k v_0 / \omega_{{\rm p}e} < 0.35$. 
As shown in Figs.~\ref{fig:EITI_noNe_noTi}(b) and \ref{fig:EITI_noNe_noTi}(d), when no interactions occur among these modes, i.e., at $k v_0 / \omega_{{\rm p}e} > 4$, we find no imaginary roots and eight real roots; the slow and fast modes of the reflected protons ($\omega = k v_{\rm dP-ref}  \pm \omega_{\rm pP-ref}$, where $+$ and $-$ correspond to the fast and slow modes, respectively), the expanding protons ($\omega = k v_{\rm dP-exp}  \pm \omega_{\rm pP-exp}$), the expanding C-ions ($\omega = k v_{\rm dC}  \pm \omega_{\rm pC}$), and Cl-ions ($\omega =  \pm \omega_{\rm pCl}$, where  $v_{\rm dCl}=0$).
The plasma frequencies ($\omega_{{\rm p}s}$), drift velocities ($v_s$), and relative drift velocities to Cl$^{15+}$ ions ($v_{{\rm d}s}$) are summarized in Table \ref{tab:table1}.
 We used $\omega_{{\rm p}e}$ value shown in Table. 1 for the normalization.

Two-stream instabilities become unstable when the slow and fast modes interact with each other, and two complex roots appear. In other words, we have solutions of $\omega = {\rm Re} [\omega ] \pm i {\rm Im} [\omega ]$.
This is clearly shown in Fig.~\ref{fig:EITI_noNe_noTi} where three unstable roots with the maximum growth rate $\gamma_m$ at $k_m v_0 / \omega_{{\rm p}e} \simeq$ 0.085, 0.17, and 2.2 are excited.
Here, the $k_m$ represents the $k$ value at the maximum growth rate $\gamma_m$.
They are unstable slow modes excited between the slow reflected-proton (Ref-P) and fast expanding-proton (Exp-P) modes, the slow Exp-P and fast expanding-C-ion (Exp-C) modes, and the slow Exp-C and fast Cl-ion (Cl) modes from the small to large $k$, and we call these three unstable modes as Ref-P, Exp-P, and Exp-C modes, respectively.
%
 \begin{figure}
  \includegraphics[width=0.49\textwidth]{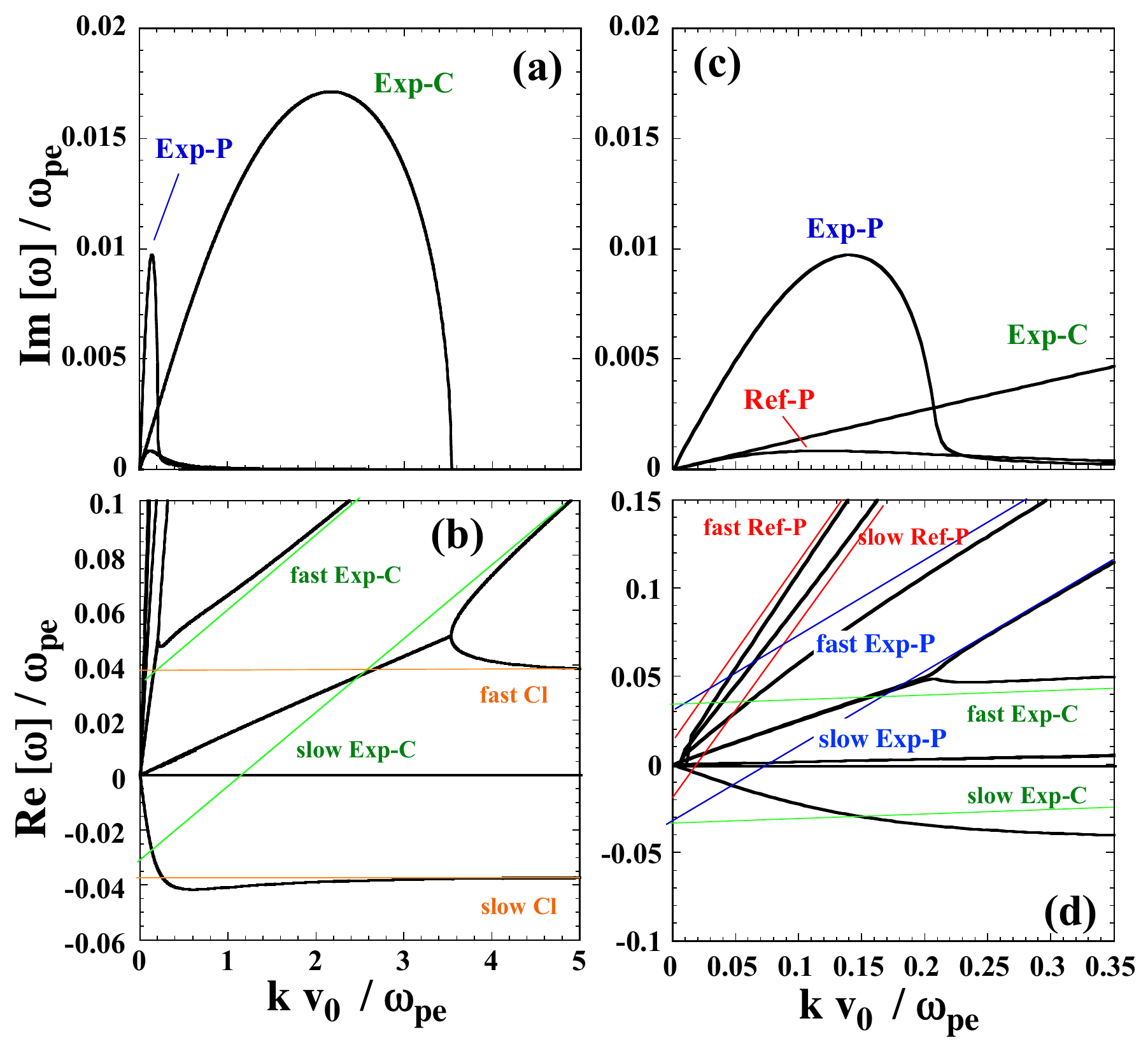}
  \caption{Results of the linear analysis of the electrostatic two-stream instability for a $\rm C_2H_3Cl$ plasma when $n_e = 7.0 \times 10^{20}$ cm$^{-3}$ and  $T_i = 2 \times 10^{-5}$ MeV. (a), (c) The imaginary part (Im [$\omega]/ \omega_{{\rm p}e}$) and (b), (d) real part (Re [$\omega]/ \omega_{{\rm p}e}$) of the normalized frequencies versus normalized wavenumber in the $x$-direction ($k v_0 / \omega_{{\rm p}e}$.  (a), (b) [(c), (d)] are plotted for $k v_0 / \omega_{{\rm p}e} < 5$ [$k v_0 / \omega_{{\rm p}e} <0.35$]. The thin red, blue, green, and orange lines in (b) and (d) are the fast and slow Ref-P, Exp-P, Exp-C, and Cl modes, respectively.}
   \label{fig:EITI_wNe_noTi}
\end{figure}
%
 \begin{figure}
  \includegraphics[width=0.49\textwidth]{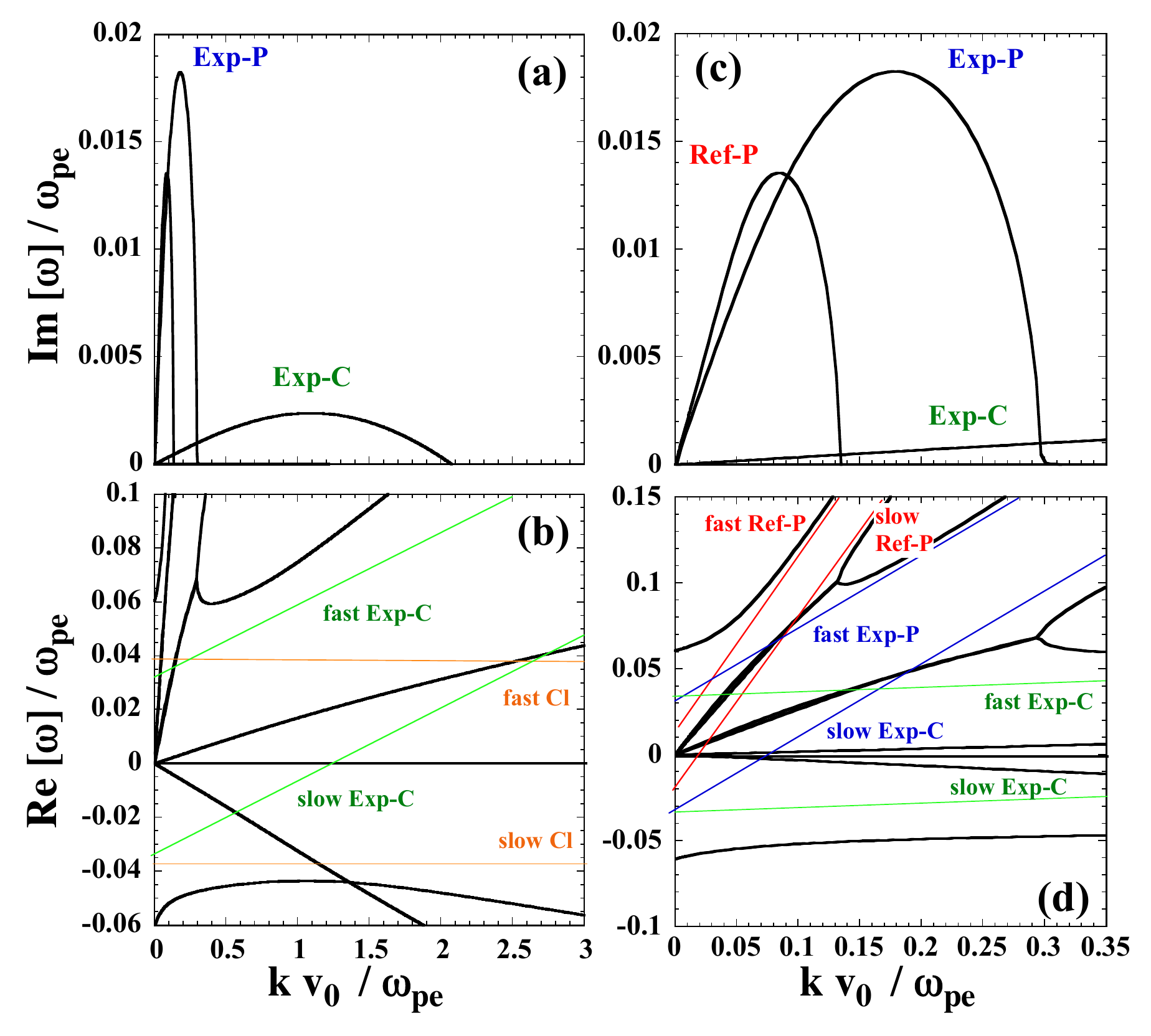}
  \caption{Results of the linear analysis of the electrostatic two-stream instability for a $\rm C_2H_3Cl$ plasma when $n_e = 0$ and $T_i = 0.02$ MeV. (a), (c) The imaginary part (Im [$\omega]/ \omega_{{\rm p}e}$) and (b), (d) real part  (Re [$\omega]/ \omega_{{\rm p}e}$) of normalized frequency versus the normalized wavenumber in the $x$-direction ($k v_0 / \omega_{{\rm p}e}$).  (a), (b) [(c), (d)] are plotted for $k v_0 / \omega_{{\rm p}e} < 3$ [$k v_0 / \omega_{{\rm p}e} <0.35$]. The thin red, blue, green, and orange lines in (b) and (d) are the fast and slow Ref-P, Exp-P, Exp-C, and Cl modes, respectively.}
   \label{fig:EITI_noNe_wTi}
\end{figure}

To clarify that the interaction between the fast and slow modes mentioned above occurs and that the slow modes are destabilized, we have carried out the linear analysis of the electrostatic two-stream instability for a $\rm C_2H_3Cl$ plasma by artificially removing some ion species.
Figure \ref{fig:EITI_iondep} shows the imaginary and real parts of instability frequency for $n_e = 0$ and  $T_i = 2 \times 10^{-5}$ MeV, which are the same parameters as Fig.~\ref{fig:EITI_noNe_noTi}, but now by neglecting one of the ion species.
Note that in all the cases, only two unstable modes are excited.

When reflected protons are removed ($n_{\rm P-ref} = 0$) [Figs.~\ref{fig:EITI_iondep}(a) - \ref{fig:EITI_iondep}(d)], the unstable modes with $\gamma_m$ at $k_m v_0 / \omega_{{\rm p}e} \simeq$ 0.17 (Exp-P mode) and 2.2 (Exp-C mode) remain and are identical to that with the reflected protons (Fig.~\ref{fig:EITI_noNe_noTi}), whereas the mode at $k_m v_0 / \omega_{{\rm p}e} \simeq$ 0.085 (Ref-P mode) disappears.
This is because the slow Ref-P mode is removed and no interaction between the fast Exp-P mode occurs.

Next, when expanding protons are removed ($n_{\rm P-exp} = 0$) [Figs.~\ref{fig:EITI_iondep}(e) - \ref{fig:EITI_iondep}(h)], whereas the unstable mode at $k_m v_0 / \omega_{{\rm p}e} \simeq$ 2.2 (Exp-C mode) remains and is identical to that with the expanding protons (Fig.~\ref{fig:EITI_noNe_noTi}), the unstable mode at $k_m v_0 / \omega_{{\rm p}e} \simeq$ 0.17 (Exp-P mode) disappears and the mode at $k_m v_0 / \omega_{{\rm p}e} \simeq 0.085$ (Ref-P mode) down-shifts to $k_m v_0 / \omega_{{\rm p}e} \simeq$ 0.06.
This is because the slow Exp-P mode is removed and no interaction between the fast Exp-C mode occurs.
As a result, the Exp-P mode is not excited.
Furthermore, since the fast Exp-P mode is removed, the slow Ref-P mode interacts with the fast Exp-C mode, and excite Ref-P mode at a down-shifted wavelength of $k_m v_0 / \omega_{{\rm p}e} \simeq$ 0.06.

Finally, when expanding C-ions are removed ($n_{\rm C} = 0$) [Figs.~\ref{fig:EITI_iondep}(i) - \ref{fig:EITI_iondep}(l)], the unstable mode at $k_m v_0 / \omega_{{\rm p}e} \simeq$ 0.085 (Ref-P mode) is identical to that with the expanding-C-ions (Fig.~\ref{fig:EITI_noNe_noTi}), the mode at $k_m v_0 / \omega_{{\rm p}e} \simeq$ 2.2 (Exp-C mode) disappears and the mode at $k_m v_0 / \omega_{{\rm p}e} \simeq$ 0.17 (Exp-P mode) down-shifts to $k_m v_0 / \omega_{{\rm p}e} \simeq$ 0.14.
This is because Exp-C mode is not excited, and since the slow Exp-P mode interacts with the fast Cl mode, Exp-P mode is excited at a down-shifted wavelength of $k_m v_0 / \omega_{{\rm p}e} \simeq$ 0.14.

From these results, we conclude that three unstable roots at $k_m v_0 / \omega_{{\rm p}e} \simeq$ 0.085, 0.17, and 2.2 for $n_e = 0$ and $T_i = 2 \times 10^{-5}$ MeV are unstable modes of Ref-P mode [reflected-proton IBTI], Exp-P mode [heavy-ion IBTI], and Exp-C mode [heavy-ion IBTI], respectively, as shown in Fig.~\ref{fig:EITI_noNe_noTi}.
%
\subsection{Cold ions with hot electrons: excitation of  the electron-ion acoustic instability (electron-ion AI) and the  ion-ion acoustic instability (ion-ion AI)}

Now we consider the effect of the electrons with $n_e = 7.0 \times 10^{20}$ cm$^{-3}$ [$1/ (k^2 \lambda_{De}^2) \neq 0$] and $T_e = 2$ MeV in addition to all four species of cold ions ($T_i = 2 \times 10^{-5}$ MeV). 
By adding electrons with Maxwellian velocity-distribution function to the cold drifting-ions, the system is possibly unstable to the electron-ion acoustic instability (electron-ion AI) as well as the ion-ion acoustic instability (ion-ion AI) \cite{Forslund1970,Akimoto1986,Karimabadi1991,Wahlund1992,Ohira2008,Kato2010b,Ross2013,Zhang2018,Jiao2019}, addition to the ion-beam two-stream instability (IBTI) \cite{Ohira2008}.
The electron-ion AI and ion-ion AI are excited in the electron background.
Whereas the electron-ion AI is excited with the relative drift between electrons and ions, the ion-ion AI is excited when there is a relative drift between ion species.
Therefore, in the multi-species ion plasma with relative drifts between ion species, both the electron-ion AI and ion-ion AI can be excited in addition to IBTI.

The instability condition for the ion-ion AI is expressed as $v_{{\rm d}s} {\rm cos} \theta \le 2c_s$, where $v_{{\rm d}s}$ is relative ion drift velocities among ion species, $c_s$ is the ion-acoustic velocity, and $\theta$ is the angle between the propagation direction of the ion-ion AI and the $x$-direction \cite{Akimoto1986,Karimabadi1991}.
Therefore, when  $v_{{\rm d}s} / 2c_s \le 1$, $\theta =0$ and the ion-ion AI occurs for $k = k_x$ and $k_y=0$, which applies to our PIC results.

When hot electrons are added, ten solutions appear.
Two new modes result from Langmuir waves, $\omega^2 = \omega_{{\rm p}e}^2  + 3 k ^2 T_e / m_e$.
These modes, which are obtained from Eq.~(\ref{eq:cold_ion_w}), are high-frequency solutions compared with other eight solutions and are not discussed further other than to state that this is equivalent to approximating Eq.~(\ref{eq:disp}) as Eqs.~(\ref{eq:cold_ion}) and (\ref{eq:cold_ion2}).

Figure \ref{fig:EITI_wNe_noTi} shows the imaginary and real parts of instability frequency for $n_e = 7.0 \times 10^{20}$ cm$^{-3}$ and $T_i = 2 \times 10^{-5}$ MeV.
We use the relativistic plasma frequency $\omega_{{\rm p}e}$ = $ \sqrt {n_e e^2 / \epsilon_0 \gamma m_e}$, with a Lorentz factor of $\gamma =   3 T_e / m_e c^2 =12$.
Note that three unstable roots appear at $k_m v_0 / \omega_{{\rm p}e} \simeq$ 0.12 (Ref-P mode), 0.14 (Exp-P mode), and 2.2 (Exp-C mode).
Compared to the $n_e =0$ case shown in Fig.~\ref{fig:EITI_noNe_noTi}, the $k_m$ value and amplitude of $\gamma_m$ are identical for Exp-C mode.
Furthermore, for Exp-P mode, $k_m$ and $\gamma_m$ are slightly smaller (by factors of 1.2 and 2.0, respectively) than the $n_e = 0$ case.  
However, for Ref-P mode, whereas $k_m$ is slightly larger (by a factor of 1.4), $\gamma_m$ is much smaller (by a factor of 17) than the $n_e = 0$ case.
This large reduction in $\gamma_m$ for Ref-P mode suggests that this mode is either the electron-ion AI or ion-ion AI.
We explain the details below.

When hot electrons are added to the cold ion dispersion relation, the $1/(k^2 \lambda_{De}^2)$ term in Eq.~(\ref{eq:cold_ion}) is important.
The real part of the dispersion relation is identical to that derived from Eq.~(\ref{eq:no_e}) when $k \to \infty$ or when $1/(k^2 \lambda_{De}^2)=0$.
However, when $k \to 0$, the fast and slow modes become Re [$\omega]/ \omega_{{\rm p}e} \to 0$ instead of being $\pm \omega_{\rm ds}$.
In our calculation, the Debye length is $\lambda_{De} = 4.0 \times 10^{-7}$ m/s and $k \lambda_{De}>1$ is satisfied when $k v_0 / \omega_{{\rm p}e} > 0.19$. For Ref-P and Exp-P modes, $\gamma_m$ occurs at $k_m v_0 / \omega_{{\rm p}e} \simeq$ 0.12 and 0.14, respectively, and the electrons are non-negligible.
The real parts of the unstable modes shown in Figs.~\ref{fig:EITI_wNe_noTi}(b) and \ref{fig:EITI_wNe_noTi}(d) reveal that whereas $\gamma_m$ for Exp-P mode occurs roughly at the interaction point of the fast Exp-C and the slow Exp-P modes, that for Ref-P mode occurs on the slow reflected-proton ion-acoustic mode ($\omega = k v_{d{\rm P-ref}}  -  \sqrt{n_{\rm P-ref} / n_e} \sqrt{k^2 / (1+ k^2 \lambda_{De}^2)} \sqrt{T_e / m_p}$) [see Eq.~(\ref{eq:cold_ion2})].

These results indicate that Ref-P mode is either the electron-ion AI or ion-ion AI; 
whereas IBTI is dominant for Exp-P mode, small modification in the real and imaginary parts of the instability frequency appears due to the electron effect;
and Exp-C mode is IBTI, since the condition for the excitation of IBTI  $k_m \lambda_{De} \gg 1$ is satisfied.
Further discussion is given below.
%
\subsection{Finite-temperature ions}
To extend the linear analysis to warm ions and hot electrons we solve Eq.~(\ref{eq:disp}) numerically with $T_i =0.02$ MeV for all ion species, this includes the reflected protons, expanding protons, C$^{6+}$ ions, and Cl$^{15+}$ ions.
\subsubsection{$T_i = 0.02$ MeV and $n_e = 0$: ion Landau damping effect}
Figure \ref{fig:EITI_noNe_wTi} shows the imaginary and real parts of the instabilities versus normalized wavenumber when  $1/(k^2 \lambda_{De}^2)=0$ or $n_e = 0$ and $T_i = 0.02$ MeV.  Three unstable solutions, Ref-P, Exp-P, and Exp-C modes, appear.
The $k_m$ and $\gamma_m$ result for Ref-P and Exp-P are similar to cold-ion case (Fig.~\ref{fig:EITI_noNe_noTi}) discussed in Part.~III-A.
However, for Exp-C mode, $\gamma_m$ is a factor of 7.2 smaller and $k_m v_0 / \omega_{{\rm p}e}$ is downshifted by a factor of 2.0.

The reduction in $\gamma_m$ and $k_m$ in Exp-C mode results from the ion Landau damping.
When $T_i = 2 \times 10^{-5}$ MeV as shown in Fig.~\ref{fig:EITI_noNe_noTi}, the $\gamma_m$ appears at the real part of $\omega$ ($\omega_m$) and $k_m$ below the resonance condition between the slow Exp-C mode ($\omega = k v_{\rm dC} - \omega_{\rm pC}$) and the fast Cl mode ($\omega =  \omega_{\rm pCl}$), that is, $\omega_m <  \omega_{\rm pCl}$ and $k_m < (\omega_{\rm pC} + \omega_{\rm pCl}) / v_{\rm dC}$.
When $T_i = 0.02$ MeV, i.e., $T_i  > 0$ as shown in Fig.~\ref{fig:EITI_noNe_wTi}, the larger-$k$ part of the instability is stabilized by the ion Landau damping in the region where $k \lambda_{D \rm{C}} \ge 1$ is satisfied \cite{Ohira2008}. Here, $\lambda_{D \rm{C}}$ is the ion Debye length for C$^{6+}$ ions.
For $T_i =0.02$ MeV, the condition of $k \lambda_{\rm DC} =1$ corresponds to $kv_{\rm 0}/\omega_{\rm pe} = 2.9$.
Since the smaller-$k$ ($k < 1/ \lambda_{D \rm{C}}$) part of the instability is still unstable, both $k_m$ and $\gamma_m$ are reduced.
%

\subsubsection{$T_i = 0.02$ MeV and $n_e = 7.0 \times 10^{20}$ cm$^{-3}$: stabilization of the reflected-proton mode}
\begin{figure}
  \includegraphics[width=0.49\textwidth]{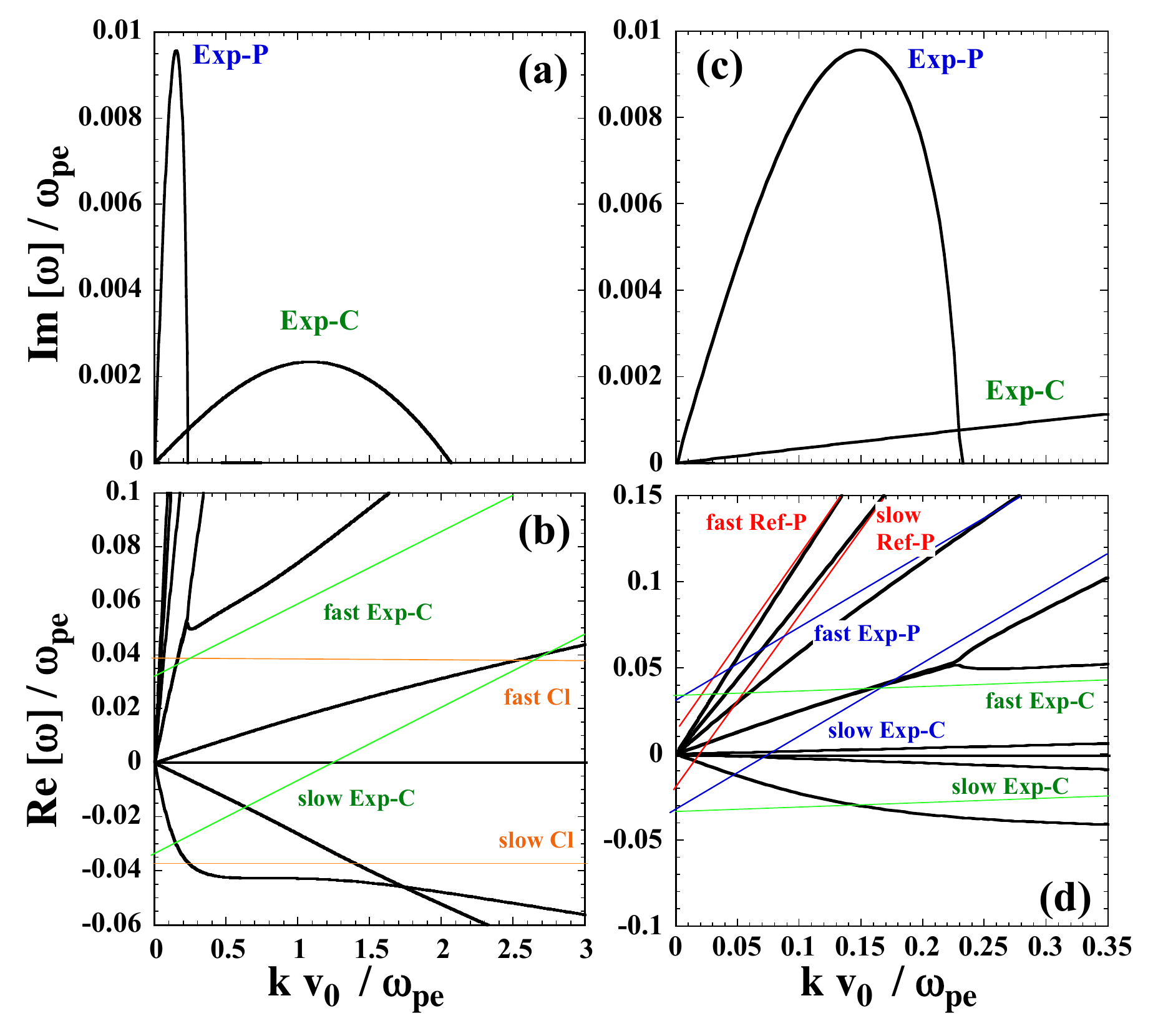}
  \caption{Results of the linear analysis of the electrostatic two-stream instability for a $\rm C_2H_3Cl$ plasma when $n_e = 7.0 \times 10^{20}$ cm$^{-3}$ and $T_i = 0.02$ MeV.
(a), (c) The imaginary part (Im [$\omega]/ \omega_{{\rm p}e}$) and (b), (d) real part  (Re [$\omega]/ \omega_{{\rm p}e}$) of normalized frequency versus the normalized wavenumber in the $x$-direction ($k v_0 / \omega_{{\rm p}e}$).  (a), (b) [(c), (d)] are plotted for $k v_0 / \omega_{{\rm p}e} < 3$ [$k v_0 / \omega_{{\rm p}e} <0.35$]. The thin red, blue, green, and orange lines in (b) and (d) are the fast and slow Ref-P, Exp-P, Exp-C, and Cl modes, respectively.
}
   \label{fig:EITI_wNe_Ti03}
\end{figure}

Figure \ref{fig:EITI_wNe_Ti03} shows the imaginary and real parts of the instability frequency versus normalized wavenumber, when $n_e = 7.0 \times 10^{20}$ cm$^{-3}$ and $T_i = 0.02$ MeV.
Note that only two unstable roots appear at $k_m v_0 / \omega_{{\rm p}e} \simeq$ 0.15, Exp-P mode, and $k_m v_0 / \omega_{{\rm p}e} \simeq$ 1.1, Exp-C mode, and Ref-P mode disappears.
In comparison with cold-ion and hot-electron case (Fig.~\ref{fig:EITI_wNe_noTi}) discussed in Part.~III-B, we find that whereas Exp-P mode is nearly identical, the maximum growth rate $\gamma_m$ for Exp-C mode is reduced by a factor of 7.2 at a downshifted wavelength, $k_m v_0 / \omega_{{\rm p}e}$ by a factor of 2.0.
These $T_i$ effects are similar to the $n_e = 0$ case.

The reduction in $\gamma_m$ occurs for Exp-C mode when $T_i=0.02$ MeV even for the $n_e=0$ case (Fig.~\ref{fig:EITI_noNe_wTi}). This implies the ion Landau damping is important.
On the other hand, the stabilization of Ref-P mode is shown in Fig.~\ref{fig:EITI_wNe_Ti03} when $T_i=0.02$ MeV and electrons are included. 
Again, this suggests that Ref-P mode is either electron-ion AI or ion-ion AI because the stability condition of the two instabilities is sensitive to the ion temperature.
%
%

%
\subsection{Identification of the instabilities}
%
%
 \begin{figure}
  \includegraphics[width=0.49\textwidth]{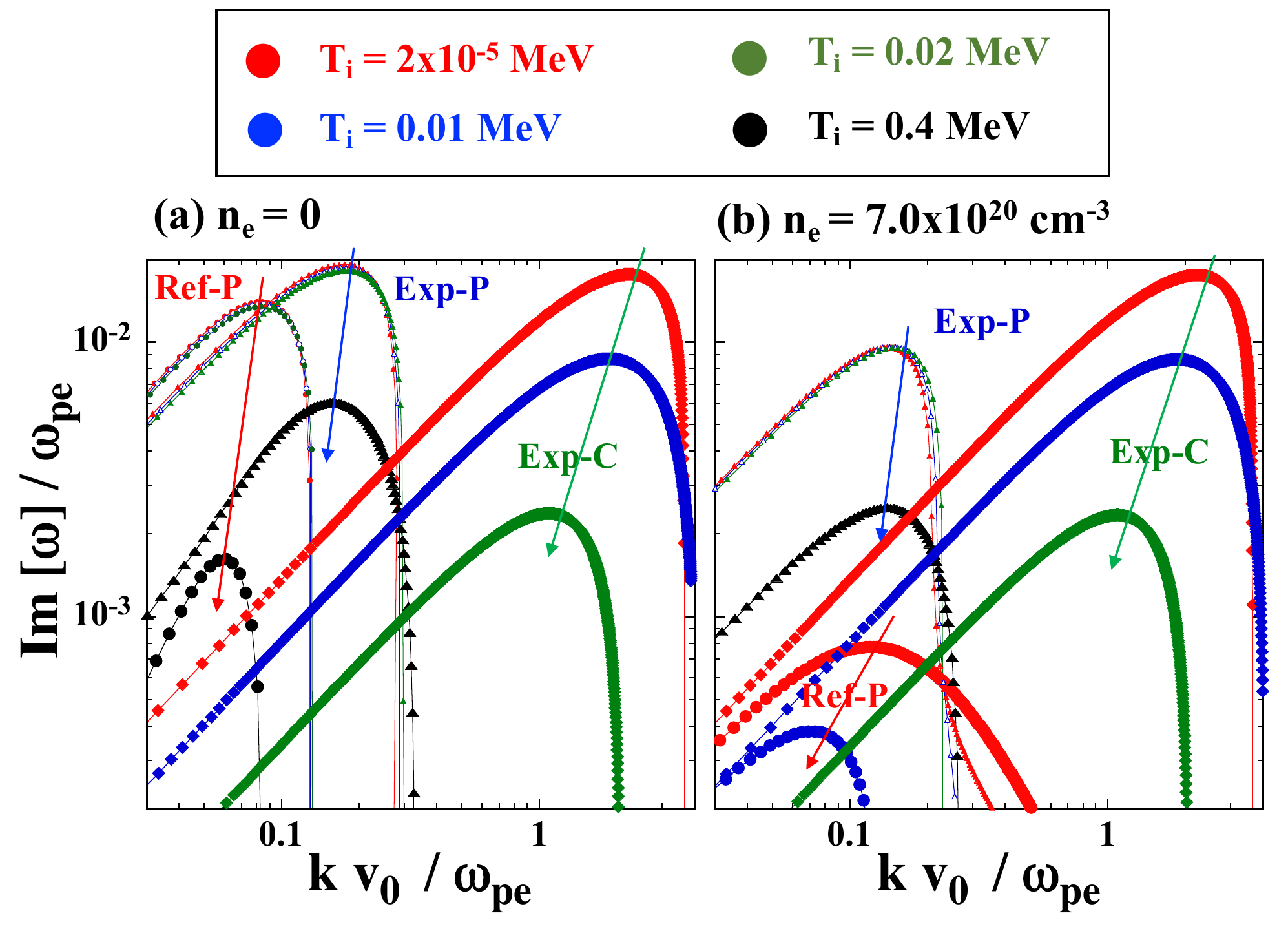}
  \caption{Results of the linear analysis of the electrostatic two-stream instability for a $\rm C_2H_3Cl$ plasma including expanding $\rm C^{6+}$ ions, $\rm Cl^{15+}$ ions, protons, and reflected protons. The normalized growth rate (Im [$\omega]/ \omega_{{\rm p}e}$) versus the wavenumber in the $x$-direction ($k v_0 / \omega_{{\rm p}e}$).
Ion temperature effect of the unstable modes for (a) $n_e = 0$ and  (b) $n_e = 7.0 \times 10^{20}$ cm$^{-3}$ for $T_i = 2 \times 10^{-5}$ MeV (red marks), $T_i = 0.01$ MeV (blue marks), $T_i = 0.02$ MeV (green marks), and $T_i = 0.4$ MeV (black marks). Ref-P mode (small-$k$ roots, filled circles), Exp-P mode (medium-$k$ roots, filled triangles), and Exp-C mode  (large-$k$ roots, filled diamonds) are shown. Arrows are guide to eyes.}
   \label{fig:EITI_Tidep}
\end{figure}

Figure \ref{fig:EITI_Tidep}(a) shows the variation of the normalized growth rate versus the normalized wavenumber by changing $T_i$ for $n_e = 0$. When $T_i = 2 \times 10^{-5}$ MeV [red marks in Fig.~\ref{fig:EITI_Tidep}(a)], which is the same parameters as Fig.~\ref{fig:EITI_noNe_noTi}, three unstable modes, Ref-P, Exp-P, and Exp-C modes  at $k_m v_0 / \omega_{{\rm p}e} \simeq$ 0.085, 0.17, and 2.17, respectively, are seen.
When $T_i$ = 0.01 and 0.02 MeV, Ref-P and Exp-P modes are similar, however, for Exp-C mode $k_m$ and $\gamma_m$ become smaller.
This results from the ion Landau damping when $k \lambda_{Di} \ge 1$ is satisfied, which corresponds to $k v_0 / \omega_{{\rm p}e} \ge 4.0$ and 2.9, respectively, for $T_i$ = 0.01 and 0.02 MeV.
Therefore, the larger-$k$ part of the unstable Exp-C mode is stabilized by the ion Landau damping when $T_i$ is increased.
Figure \ref{fig:EITI_Tidep}(b) displays the variation of the normalized growth rate versus the normalized wavenumber for 4 values of $T_i$ and $n_e = 7.0 \times 10^{20}$ cm$^{-3}$. 

Figure 5(b) in Paper I and Fig.~2(c) illustrate how the $T_i$ are inferred from the PIC calculations of the velocity distributions. At 4 ps, these are $\sim$0.3, $\sim$0.1, and $\sim$0.1 MeV for the expanding protons, C$^{6+}$ and Cl$^{15+}$ ions, respectively. In Fig.~\ref{fig:EITI_Tidep}, results for $T_i  = 0.4$ MeV are shown with black marks. A large reduction in $\gamma_m$ for Ref-P and Exp-P modes when $n_e = 0$ [Fig.~\ref{fig:EITI_Tidep}(a)] and  for Exp-P mode when $n_e = 7.0 \times 10^{20}$ cm$^{-3}$ [Fig.~\ref{fig:EITI_Tidep}(b)] results from  the ion Landau damping. Exp-C mode is also stabilized via the ion Landau damping from a lower temperature of $T_i \geq 0.03$ MeV and there are no unstable roots.

By comparing Figs.~\ref{fig:EITI_Tidep}(a) and \ref{fig:EITI_Tidep}(b), we find the following effects of including hot electrons.
(A) For Ref-P mode, the maximum growth rate $\gamma_m$ is reduced more than an order of magnitude at slightly (by a factor of 1.4) up-shifted $k_m v_0 / \omega_{{\rm p}e}$.
The upshift depends upon $T_i$; this mode is stabilized ($\gamma_m / \omega_{{\rm p}e} \simeq 6 \times 10^{-6}$) when $T_i = 0.02$ MeV  as shown in Fig.~\ref{fig:EITI_wNe_Ti03}. Ref-P mode is either the electron-ion AI or ion-ion AI.
(B) For Exp-P mode, $\gamma_m$ is reduced by a factor of 2 when including hot electrons, and independent of $T_i$ at $T_i < 0.1$ MeV. Exp-P mode is IBTI.
(C) For Exp-C mode, $\gamma_m$  is independent of the hot electrons and strongly depends on $T_i$, so Exp-C mode is IBTI.

For Ref-P mode, the observed reduction of $\gamma_m$ and upshift of $k_{\rm m}$, which are observed by including hot-electrons, suggest that this mode is either the ion-ion AI or electron-ion AI.
For the ion-ion AI, when the relative drift velocity between the different populations of ions is larger than the ion thermal velocity $v_{{\rm th}i}$, $T_i=0$ can be assumed.
When $T_i = 0.02$ MeV, the relative drift velocity between the reflected and expanding protons  ($v_{\rm P-ref} - v_{\rm P-exp} = 6.4 \times 10^{-2} c$) is larger than the thermal velocity of protons ($v_{\rm thP} = 6.5 \times 10^{-3}c$).
This implies that $T_i$ effects are negligible for Ref-P mode so that this is not likely to be the ion-ion AI.
In contrast, the growth rate of the electron-ion AI has $T_i$ dependence as a result of the ion Landau damping \cite{Ichimaru1992}. Therefore, we conclude that Ref-P mode is the electron-ion AI.
\section{Discussion}
%
  \begin{figure}[t]
  \includegraphics[width=0.45\textwidth]{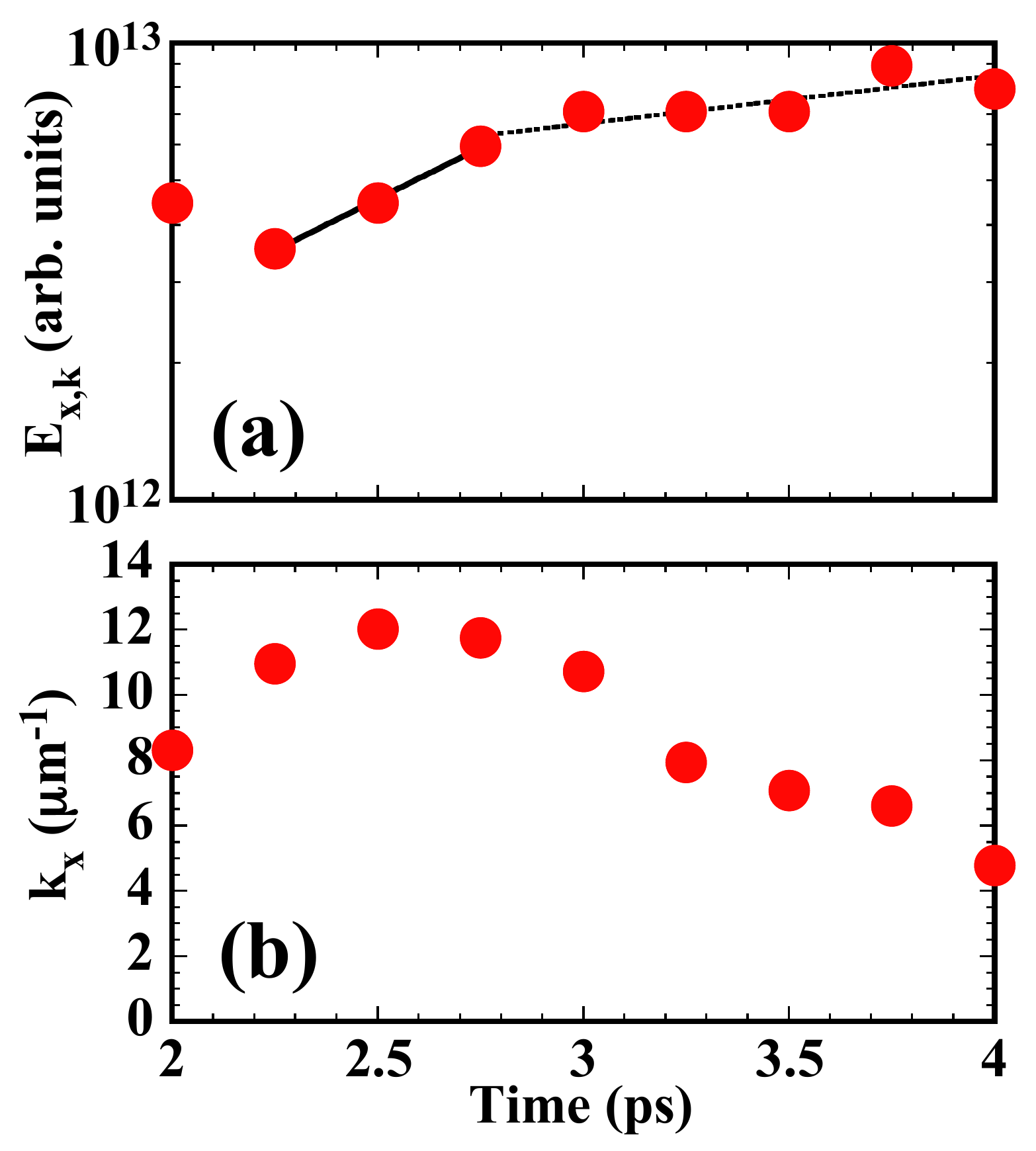}
  \caption{The temporal evolution of (a)  the peak amplitude of the electrostatic fluctuation $E_{x,k}$ and (b) the dominant $k_x$ in the PIC from the power spectrum of $E_x$ taken at the width of $\Delta x$ = 10 $\mu$m in a few $\mu$m upstream region of a shock in a $\rm C_2H_3Cl$ plasma. Solid and dotted lines in (a) represent growth rates of $E_x$, $\gamma = 1.0 \times 10^{12}$ s$^{-1}$ and $2.4 \times 10^{11}$ s$^{-1}$, derived from the exponential fits to the data at $t = 2.25 - 2.75$ ps and $2.75 - 4.0$ ps, respectively.}
   \label{fig:R1}
\end{figure}

From the results shown in Sec.~III, we conclude that the most unstable mode with the largest growth rate is Exp-P mode, which is IBTI with a small modification in both the real and imaginary parts of the frequency due to the hot electrons.
Ref-P mode is the electron-ion AI ($k \lambda_{De} < 1$), whose growth rate strongly depends on $T_i$.
Exp-C mode is IBTI ($k_m \lambda_{De} \gg 1$), which is stable because of the large ion Landau damping.

Now, we compare the results of the linear analysis with the PIC simulations. 
Figures \ref{fig:R1}(a) and \ref{fig:R1}(b) represent the temporal variation of the peak amplitude of the electrostatic fluctuation $E_{x,k}$ and the dominant $k_x$, respectively, obtained from the PIC simulations, where $E_{x,k} = (1/\sqrt{2\pi}) \int E_x(x) e^{-ik_x x} dx$ is the Fourier component of electric field.
These values are derived from the power spectrum $E_{x,k}^2$ versus $k_x$, as shown in Figs.~4(d) and 4(e) of Paper I.
We find that $E_x$ shows the fast growth early in time at $t = 2.25 - 2.75$ ps and the slower growth later in time at $t = 2.75 - 4.0$ ps. 
The growth rate $\gamma^{\rm PIC}$ and the dominant $k_x^{\rm PIC}$ values at 4.0 ps are $\gamma^{\rm PIC} = 2.4 \times 10^{11}$ s$^{-1}$  and  $k_x^{\rm PIC} = 4.8$ $\mu$m$^{-1}$, respectively.

As described in Sec.~III-D, proton temperature $T_{\rm P}$ derived from the velocity spread obtained by PIC at 4 ps is $T_{\rm P} \sim 0.3$ MeV. The maximum growth rate ($\gamma_m^{\rm Exp-P}$) and the $k_x$ value at the maximum growth rate ($k_m^{\rm Exp-P}$) for the most unstable mode, Exp-P mode, derived from the linear analysis for $T_i = 0.4$ MeV shown in Fig.~9(b) are $\gamma_m^{\rm Exp-P} = 1.1 \times 10^{12}$ s$^{-1}$ ($\gamma_m^{\rm Exp-P}  / \omega_{{\rm p}e} =  2.5 \times 10^{-3}$) and  $k_m^{\rm Exp-P} = 1.8$ $\mu$m$^{-1}$ ($k_m^{\rm Exp-P}  v_0 / \omega_{{\rm p}e} = 0.14$), respectively.
We find that $k_m^{\rm Exp-P}$ and $k_x^{\rm PIC}$ ($k_x^{\rm PIC}  v_0 / \omega_{{\rm p}e} = 0.37$) agree relatively well with each other, while $\gamma_m^{\rm Exp-P}$ is more than a factor of 4  larger than $\gamma^{\rm PIC}$ ($\gamma^{\rm PIC}  / \omega_{{\rm p}e} =  5.6 \times 10^{-4}$).

\begin{figure}[t]
  \includegraphics[width=0.45\textwidth]{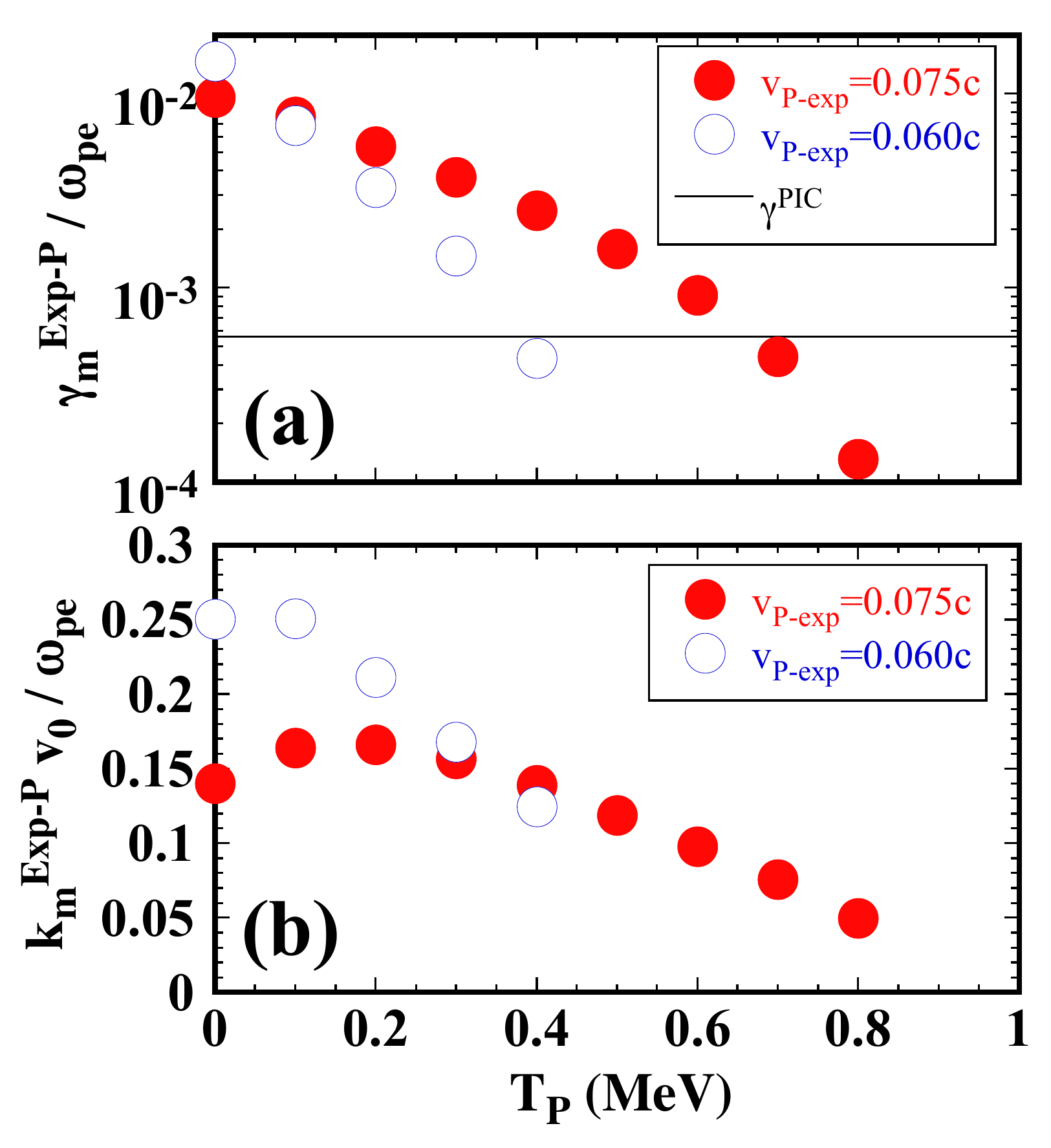}
  \caption{Proton temperature $T_{\rm P}$ dependence of (a) the normalized maximum growth rate ($\gamma_m^{\rm Exp-P}  / \omega_{{\rm p}e}$) and (b) the normalized wavenumber in the $x$-direction ($k_m^{\rm Exp-P}  v_0 / \omega_{{\rm p}e}$) of Exp-P mode obtained from the linear analysis for $n_e = 7.0 \times 10^{20}$ cm$^{-3}$ at 4.0 ps. The drift velocity of the expanding protons $v_{\rm P-exp}$ used in the linear analysis is $0.075c$ (filled circles) and $0.060c$ (open circles).
 The normalized growth rate ($\gamma^{\rm PIC}  / \omega_{{\rm p}e}$) and the wavenumber in the $x$-direction ($k_x^{\rm PIC}  v_0 / \omega_{{\rm p}e}$) obtained from the PIC simulation at 4.0 ps are $5.6 \times 10^{-4}$ [horizontal solid line in (a)] and 0.37, respectively.}
   \label{fig:R2}
\end{figure}

One of the possible explanations for $\gamma_m^{\rm Exp-P} > \gamma^{\rm PIC}$ is the difference in the proton temperature $T_{\rm P}$.
We inferred $T_{\rm P}$ from the velocity spread of the expanding protons from the PIC simulation, where the velocity distribution is far from a simple Maxwell distribution as shown in Fig.~\ref{fig:8}. 
Therefore, it is not easy to derive an accurate $T_{\rm P}$.
Figures \ref{fig:R2}(a) and \ref{fig:R2}(b) show the normalized maximum growth rate ($\gamma_m^{\rm Exp-P} / \omega_{{\rm p}e}$) and the normalized wavenumber in the $x$-direction ($k_m^{\rm Exp-P} v_0 / \omega_{{\rm p}e}$) of Exp-P mode, respectively, as a function of the proton temperature $T_{\rm P}$ for $n_e = 7.0 \times 10^{20}$ cm$^{-3}$ at 4.0 ps. The drift velocity of the expanding protons $v_{\rm P-exp}$ is $0.075c$ (filled circles) as shown in Table \ref{table:1}. 
We see that to achieve the growth rate obtained by PIC at 4 ps, $\gamma^{\rm PIC} / \omega_{{\rm p}e} = 5.6 \times 10^{-4}$, $T_{\rm P} \simeq 0.7$ MeV is required.
This value is more than a factor of 2 larger than the $T_{\rm P}$ obtained by PIC calculations.

A second possible explanation for $\gamma_m^{\rm Exp-P} > \gamma^{\rm PIC}$ is the difference in the drift velocity of the protons $v_{\rm P-exp}$.
Table \ref{tab:table2} summarizes drift velocities of expanding protons, C$^{6+}$ and Cl$^{15+}$ ions derived from the 2D PIC simulations at $t = 4.0$ ps for a $\rm C_2H_3Cl$ plasma.
In our linear analysis, we used $v_{\rm P-exp} =0.075c$, which is the peak value of the expanding proton distribution function or the drift velocity of the high-velocity component $v_{\rm P}^{\rm H}$ as shown in Fig.~\ref{fig:8}(c).
In comparison, using a medium-velocity component $v_{\rm P}^{\rm M} = 0.06c$ for $v_{\rm P-exp}$,  there is reasonable agreement with theory.
In Figs.~\ref{fig:R2}(a) and \ref{fig:R2}(b), $\gamma_m^{\rm Exp-P} / \omega_{{\rm p}e}$ and $k_m^{\rm Exp-P}  v_0 / \omega_{{\rm p}e}$, respectively, for $v_{\rm P}^{\rm H} = 0.06c$ case are represented in open circles.
We find that $\gamma^{\rm PIC} / \omega_{{\rm p}e} = 5.5 \times 10^{-4}$ agrees well with $\gamma_m^{\rm Exp-P} / \omega_{{\rm p}e} = 4.3 \times 10^{-4}$ at $T_{\rm P} = 0.4$ MeV.
In this case $k_x^{\rm PIC}  v_0 / \omega_{{\rm p}e} = 0.36$ agree within a factor of 3 with $k_m^{\rm Exp-P} v_0 / \omega_{{\rm p}e} = 0.12$ at $T_{\rm P} = 0.4$ MeV.
These results indicate that Exp-P mode, which is an IBTI, is in the nonlinear regime at 4.0 ps. This nonlinearity occurs later in time and results in the saturation in the growth of the wave amplitude, the broadening of the proton velocity distribution, and the larger wave number compared with that for the resonant mode. We discuss this in the following.

%
%

%

We have also conducted the linear analysis using the plasma densities, drift velocities, and temperatures obtained from the PIC calculation at 2.5 ps.
The temperatures of the expanding protons, C$^{6+}$ and Cl$^{15+}$ ions are $\sim$0.07, $\sim$0.07, and $\sim$0.14 MeV, respectively.
When $T_i = 0.07$ MeV, only Exp-P mode is excited and $\gamma_m^{\rm Exp-P}  / \omega_{{\rm p}e} =  1.9 \times 10^{-3}$ and $k_m^{\rm Exp-P} v_0 / \omega_{{\rm p}e} = 0.40$ are obtained.
Compared with the growth rate and $k_x$ values derived from the PIC at 2.5 ps (shown in Fig.~10), $\gamma^{\rm PIC}  / \omega_{{\rm p}e} =  1.5 \times 10^{-3}$ ($\gamma^{\rm PIC}= 1.0 \times 10^{12}$ s$^{-1}$) and $k_x^{\rm PIC} v_0 / \omega_{{\rm p}e} = 0.51$ ($k_x^{\rm PIC} = 12$ $\mu$m$^{-1}$), those from the linear analysis agree well within a factor of 1.3.

A better agreement between the results from the PIC and linear analysis is achieved when  the plasma parameters at 2.5 ps are used compared with those at 4.0 ps.
This might result from the fact that the velocity distribution of the expanding protons shown in  Fig.~\ref{fig:8} is close to a Maxwellian distribution at early time, similar to the distributions assumed by theory.
In other words, at 2.5 ps Exp-P mode or IBTI is in the linear regime, and by 4.0 ps this instability has entered a nonlinear regime. Therefore, Exp-P mode is clearly observed in the PIC at 2.5 ps as the linear analysis predicts.

In Table \ref{tab:table2}, the expanding velocities estimated from $E_{\rm TNSA}$ at 4 ps are also shown.
These velocities are taken from Figs.~5 and 6 of Paper I.
We find that, $v_{\rm C}^{\rm L}  \simeq  v_{\rm P}^{\rm TNSA}=0.034 c$, $v_{\rm C}^{\rm H} \simeq v_{\rm P}^{\rm L} =0.042 c$, $v_{\rm P}^{\rm M}$ = 0.90 $v_{\rm P}^{\rm TNSA} =  0.060 c$, and $v_{\rm P}^{\rm H}$ = 1.1 $v_{\rm P}^{\rm TNSA} = 0.075 c$.
These results suggest that Exp-P mode, which is IBTI between expanding protons and C$^{6+}$ ions, heats C$^{6+}$ ions and generates the high-velocity component of $v_{\rm C}^{\rm H}$; at the same time, heats protons, generates the low-velocity component of $v_{\rm P}^{\rm L}$, and down-shifts $v_{\rm P}^{\rm M}$ from $v_{\rm P}^{\rm TNSA}$.
The velocity spectrum of C$^{6+}$ ions at 2.0 and 4.0 ps is shown in Fig.~5 of  Paper I, and the temporal variation of the low- and high-velocity components of C$^{6+}$ ions, $v_{\rm C}^{\rm L}$ and $v_{\rm C}^{\rm H}$ respectively, is shown in Fig. 6 of Paper I.
\begin{table}[h]
\caption{\label{tab:table2} Drift velocities of expanding protons, C$^{6+}$ and Cl$^{15+}$ ions derived from the 2D PIC simulations at $t = 4.0$ ps for a $\rm C_2H_3Cl$ plasma.
The expanding velocities estimated from $E_{\rm TNSA}$ at 4 ps are also shown.
These velocities are taken from Figs.~5 and 6 of Paper I.
}
\label{table:2}
\centering
\begin{ruledtabular}
\begin{tabular}{l l l }
\multicolumn{1}{c}{\textbf{ }} & \textbf{Definition} &    \\
\hline
Proton velocities \\
\hspace {2mm}Low-velocity component & $v_{\rm P}^{\rm L} / c$  &  0.042   \\
\hspace {2mm}Medium-velocity component &$v_{\rm P}^{\rm M}  / c$ &  0.060  \\
\hspace {2mm}High-velocity component & $v_{\rm P}^{\rm H}  / c$ &  0.075  \\
\hspace {2mm}TNSA velocity &  $v_{\rm P}^{\rm TNSA}  /c$ & 0.067   \\
C$^{6+}$-ion velocities   \\
\hspace {2mm}Low-velocity component & $v_{\rm C}^{\rm L}/ c$  &  0.033   \\
\hspace {2mm}High-velocity component & $v_{\rm C}^{\rm H} / c$ &  0.044  \\
\hspace {2mm}TNSA velocity &  $v_{\rm C}^{\rm TNSA} /c$ & 0.03   \\
Cl$^{15+}$-ion velocities   \\
\hspace {2mm}  & $v_{\rm Cl} / c$  &  0.030   \\
\hspace {2mm}TNSA velocity &   $v_{\rm Cl}^{\rm TNSA} /c$ & 0.030   \\
\hline
\end{tabular}
\end{ruledtabular}
\end{table}

Possible explanations for the up-shift of $v_{\rm P}^{\rm H}$ from $v_{\rm P}^{\rm TNSA}$ are either by Exp-P mode or Ref-P mode.
Ref-P mode, which is the electron-ion AI between electrons and reflected protons, can up-shift expanding protons and down-shift reflected protons.
However, since the calculated $\gamma_m$ for Ref-P mode is more than an order of magnitude smaller than that of Exp-P mode, the contribution of Ref-P mode should be negligible.
Therefore, we conclude the up-shift of $v_{\rm P}^{\rm H}$ from $v_{\rm P}^{\rm TNSA}$ also results from Exp-P mode.
Furthermore, $v_{\rm Cl}  = v_{\rm Cl}^{\rm TNSA}  = 0.030 c$ and no heating occurs for Cl$^{15+}$ ions.
This is consistent with the stabilization of Exp-C mode, which is BTI but a finite ion temperature causes the ion Landau damping.

To enhance the number of reflected and accelerated ions in a collisionless shock, it is important to have a large number of expanding ions with velocities above $v_L^i$. 
Excitation of electrostatic two-stream instabilities is a possible solution to increase the number of the reflected ions since ions are heated by them.
In a multicomponent C$_2$H$_3$Cl plasma, excitation of the electrostatic two-stream instability between expanding protons and C$^{6+}$ ions, which is Exp-P mode, results in heating of expanding protons and C$^{6+}$ ions, and the number of the expanding ions with velocities above $v_L^i$ increases.
Furthermore, excitation of the electrostatic two-stream instability between expanding and reflected protons, which is Ref-P mode, results in heating of the expanding and reflected protons.
The heating of expanding protons increases the number of the expanding protons with velocities above $v_L^i$, and results in a larger number of reflected protons.
As a consequence, Ref-P mode is enhanced, and the energy spread of the reflected protons increases.
Therefore, exciting Exp-P mode, while stabilizing Ref-P mode, is an ideal condition for generating a large number of quasi-monoenergetic ions.
In this study, we have highlighted that a large growth rate of Ref-P mode occurs when electrons are neglected, but a more realistic treatment including hot electrons suppresses this growth by more than an order of magnitude.
This results from the suppression of IBTI and the excitation of the low growth-rate electron-ion AI for Ref-P mode.
Furthermore, by using a multicomponent C$_2$H$_3$Cl plasma, IBTI between expanding protons and C$^{6+}$ ions is excited, and the temperature of the expanding protons increases.
This results in the ion Landau damping and further stabilization of Ref-P mode.
\section{Summary}
In summary, 2D PIC simulations are used to study the formation of the laser-driven electrostatic collisionless shock in a multicomponent C$_2$H$_3$Cl plasma.
The upstream expanding ion populations are accelerated by the non-oscillating electric field, which accelerates the heavier and lighter ions to different velocities.
Furthermore, part of the ion populations in the upstream region is reflected and accelerated at the shock.
These relative drifts between two ion populations result in the excitation of an electrostatic two-stream instability, which leads to the broadening of the upstream expanding-proton distribution.  

A linear analysis of the instabilities for a $\rm C_2H_3Cl$ plasma is carried out using the one-dimensional electrostatic plasma dispersion function for unmagnetized collisionless plasmas to identify the instability.
The most unstable mode is the expanding-proton mode, which is the electrostatic ion-beam two-stream instability excited between the expanding protons and C ions.
The reflected-proton mode, which is the electrostatic electron-ion acoustic instability excited between the reflected protons and electrons, is also unstable with the smaller growth rate compared with the expanding-proton mode, and the growth rate depends on the ion temperature.
The expanding-C-ion mode, which is the electrostatic ion-beam two-stream instability excited between the expanding C and Cl ions, is stable results from the large ion Landau damping.

In a multicomponent, near critical-density plasma, the fast-growing electrostatic ion-beam two-stream instability is excited. This increases the number of reflecting and accelerating ions at an electrostatic collisionless shock, leading to brighter quasi-monoenergetic ion beams.

\section{Acknowledgement}
We thank T. Sano for the useful discussion. This research was partially supported by Japan Society for the Promotion of Science (JSPS) KAKENHI Grant No. JP15H02154, JP17H06202, JP19H00668, JP19H01893, JSPS Core-to-Core Program B. Asia-Africa Science Platforms Grant No. JPJSCCB20190003,  EPSRC grant EP/L01663X/1 and EP/P026796/1, the joint research project of the Institute of Laser Engineering, Osaka University (2020B2-044). YO is supported by Leading Initiative for Excellent Young Researchers, MEXT, Japan.

\section{Appendix: Temporal evolution of the electron temperature and a DC electric field}
%
%
\begin{figure}
  \includegraphics[width=\linewidth]{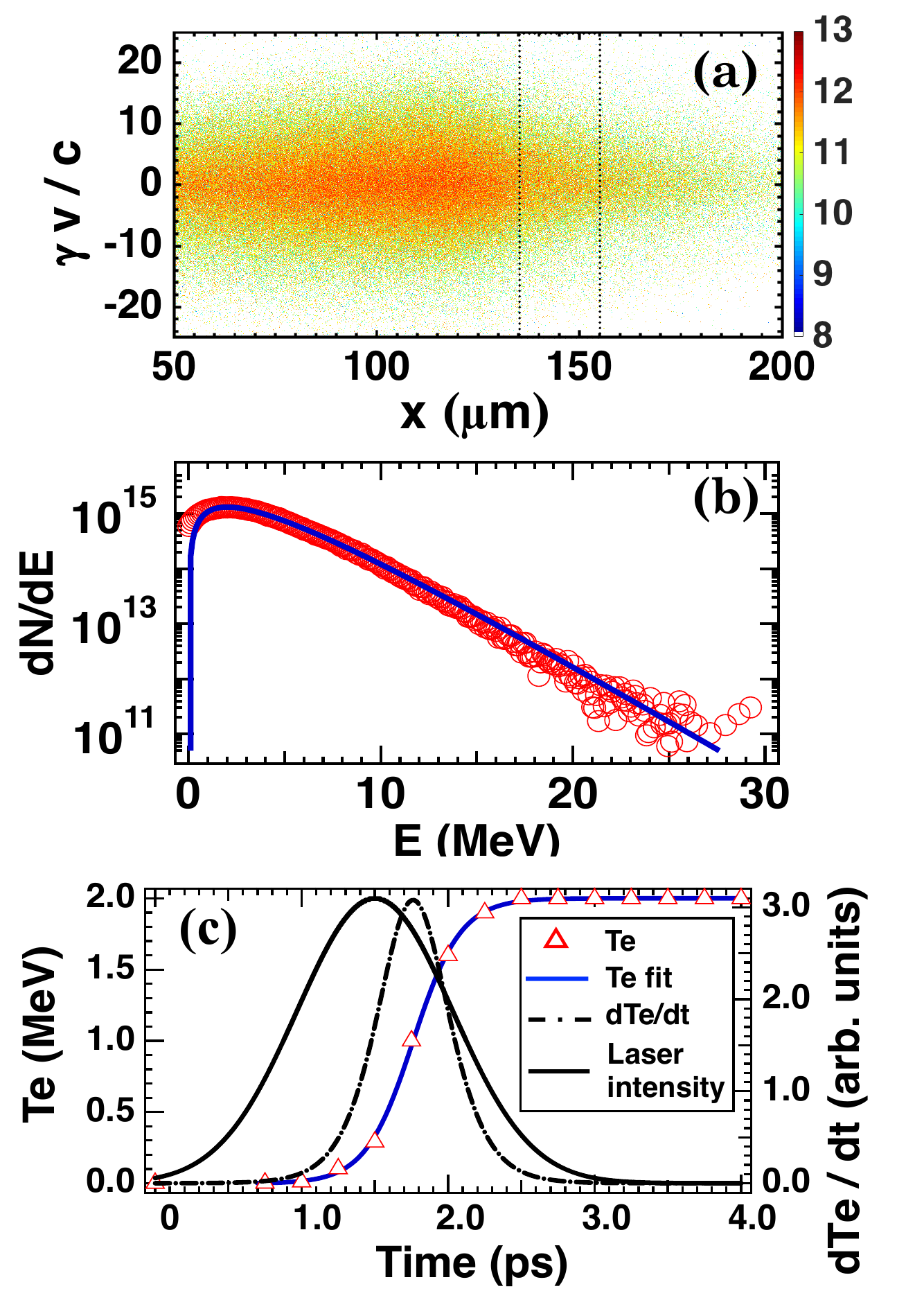}
  \caption{(a) The electron phase-space  $\gamma v / c$ versus $x$, in a $\rm C_2H_3Cl$ plasma at \textit{t} = 4.0 ps. The color scale shows the number of electrons in a log scale. (b) The electron energy spectrum taken at $\Delta x$ = 20 $\mu$m in the upstream region of the shock front in a $\rm C_2H_3Cl$ (circles) plasma at \textit{t} = 4.0 ps. The electron energy spectrum is fitted with a 2D-relativistic Maxwellian $f(E) = a  E \mathrm{exp}(-E/T_e)$ (solid line), where $a$ is constant, $E$ is the energy of electrons, $T_e$ gives the electron temperature in the upstream region. (c) The temporal evolution of $T_e$ (open triangles) calculated by fitting a 2D-relativistic Maxwellian at each time. The $T_e$ is very well fitted with a sigmoid function, $S(t) = 1/(1+e^{-at})$ (blue solid line), and the derivative ($d T_e / dt$) of $S(t)$ (dot-dashed line), which peaks at \textit{t} = 1.75 ps.  Here $a$ is a fitting constant. Normalized temporal evolution of the laser intensity (black solid line), which peaks at \textit{t} = 1.5 ps, is also shown as a reference.}
\label{fig:4}
\end{figure}

The interaction of high-intensity laser with a relativistic near critical-density plasma results in uniform electron heating via $\vec {J} \times \vec {B}$ mechanism \cite{Kruer1985}. Figure \ref{fig:4}(a) shows the electron phase-space in a $\rm C_2H_3Cl$ plasma at \textit{t} = 4.0 ps, where the vertical axis shows the four velocity $\gamma v / c$ ($\gamma$ is the Lorentz factor) in the $x$-direction. Figure \ref{fig:4}(b) represents the electron energy spectrum taken at $\Delta x$ = 20 $\mu$m in the upstream region just ahead of the shock front, this is shown by a vertical box in Fig.~\ref{fig:4}(a). To estimate the electron temperature ($T_e$) a 2D-relativistic Maxwellian $f(E) \propto E \mathrm{exp}(-E/T_e)$ is used to fit the electron energy spectrum as shown in Fig.~\ref{fig:4}(b). The extracted electron temperature is $T_e \simeq$ 2.0 MeV.  The temporal evolution of $T_e$ is represented in Fig.~\ref{fig:4}(c).
Early in time, at \textit{t} = 1.0 ps, $T_e$ is nearly equal to the initial temperature of 500 eV. After the interaction of the laser peak at $t=1.0$ ps, $T_e$ rises sharply and reached $\simeq$ 2.0 MeV at \textit{t} = 2.50 ps and remain the same throughout the simulations. 
The time evolution of $T_e$ is very well fitted with a Sigmoid function, $S(t) = 1/(1+e^{-at})$, where $a$ is a fitting constant, and the derivative of this function peaks at \textit{t} = 1.75 ps as shown in Fig.~\ref{fig:4}(c).
 \begin{figure}[b]
  \includegraphics[width=\linewidth]{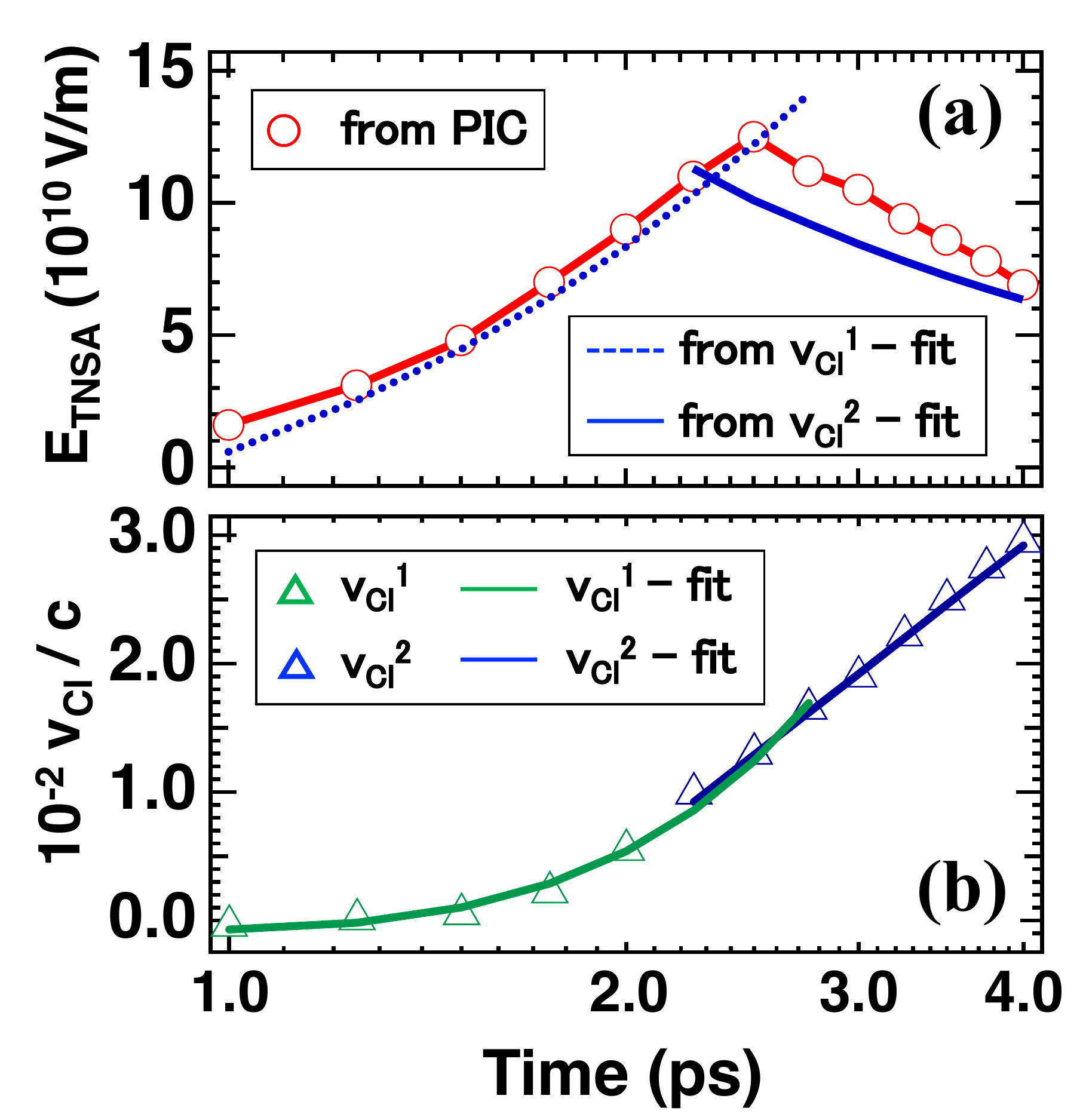}
  \caption{(a) The temporal evolution of $E_{\rm TNSA}$ measured from the PIC (open circles, red line is guide to eyes) and derived from the $v_{\rm Cl}$ (blue lines) shown in (b). (b) The velocity of Cl$^{15+}$ ions $v_{\rm Cl}$ (open triangles) derived from the peak of velocity spectrum $dN/dv_{\rm Cl}$ taken at $\Delta x$ = 3 $\mu$m in the upstream region. The $v_{\rm Cl}$ is fitted to the 2nd order polynomial from \textit{t} = 1.0 to 2.75 ps (green line), and a logarithmic curve from \textit{t} = 2.25 to 4.0 ps (blue line). In (a), $E_{\rm TNSA}$ derived from the $v_{\rm Cl}$ follows a $\propto t$ from \textit{t} = 1.0 to 2.75 ps (blue dotted line) and a 1/\textit{t} dependences from \textit{t} = 2.25 to 4.0 ps (blue solid line). Note that the $x$ axis is in the log scale.}
 \label{fig:10}
\end{figure}

The ions in the upstream expanding plasma are accelerated by a uniform sheath electric field $E_{\rm TNSA}$.
Figure \ref{fig:10}(a) shows the temporal evolution of $E_{\rm TNSA}$ measured from the PIC simulation (open circles).
$E_{\rm TNSA}$ peaks at $t \simeq 2.5$ ps.
To qualify $E_{\rm TNSA}$ obtained from PIC, we estimate it  by the temporal evolution of the expanding ion velocity in the upstream region.
Figure \ref{fig:10}(b) presents the velocity of Cl$^{15+}$ ions ($v_{\rm Cl}$, open triangles) taken at $\Delta x$ = 3 $\mu$m in the upstream region of a $\rm C_2H_3Cl$ plasma. The temporal variation of $v_{\rm Cl}$ at $t = 1.0 - 2.75$ ps shows a $t^2$ dependence on time, and a 2nd order polynomial is used to fit the $v_{\rm Cl}$ as shown in Fig.~\ref{fig:10}(b).
Later in time, $t = 2.25 - 4.0$ ps, the $v_{\rm Cl}$ follows logarithmic dependence on time, and we use it to fit $v_{\rm Cl}$.
$E_{\rm TNSA}$ can be estimated by equating the electrostatic forces ($ZeE_{\rm TNSA}$) with the acceleration ($Am_pdv/dt$) in the upstream region, where $Z$ is the charge state, $A$ is the mass number of the ion, $m_p$ is the mass of the proton, $e$ is the electron charge, and $dv/dt$ is the acceleration.
Therefore, $E_{\rm TNSA} $ is expressed as $E_{\rm TNSA} = (A/Z)(m_p/e)(dv/dt)$.
A derivative of the fitted $v_{\rm Cl}$ with $Z/A$ = 0.429 for Cl ions shows that the $E_{\rm TNSA}$ $\propto t$ at $t = 1.0 - 2.75$ ps and a 1/$t$ dependence at $t = 2.25 - 4.0$ ps as shown in Fig.~\ref{fig:10}(a).
$E_{\rm TNSA}$ obtained from the PIC simulations and estimated from $v_{\rm Cl}$ show the same trend and agree relatively well with each other. 
Therefore, it is verified that Cl$^{15+}$ ions are accelerated by $E_{\rm TNSA}$.
The logarithmic dependence of $v_{\rm Cl}$ and 1/\textit{t} dependence of $E_{\rm TNSA}$ are predicted by \citet{Mora2003}.
This temporally changing $E_{\rm TNSA}$ field accelerates the upstream ions to a uniform velocity over time.


\begin{thebibliography}{34}%
\makeatletter
\providecommand \@ifxundefined [1]{%
 \@ifx{#1\undefined}
}%
\providecommand \@ifnum [1]{%
 \ifnum #1\expandafter \@firstoftwo
 \else \expandafter \@secondoftwo
 \fi
}%
\providecommand \@ifx [1]{%
 \ifx #1\expandafter \@firstoftwo
 \else \expandafter \@secondoftwo
 \fi
}%
\providecommand \natexlab [1]{#1}%
\providecommand \enquote  [1]{``#1''}%
\providecommand \bibnamefont  [1]{#1}%
\providecommand \bibfnamefont [1]{#1}%
\providecommand \citenamefont [1]{#1}%
\providecommand \href@noop [0]{\@secondoftwo}%
\providecommand \href [0]{\begingroup \@sanitize@url \@href}%
\providecommand \@href[1]{\@@startlink{#1}\@@href}%
\providecommand \@@href[1]{\endgroup#1\@@endlink}%
\providecommand \@sanitize@url [0]{\catcode `\\12\catcode `\$12\catcode
  `\&12\catcode `\#12\catcode `\^12\catcode `\_12\catcode `\%12\relax}%
\providecommand \@@startlink[1]{}%
\providecommand \@@endlink[0]{}%
\providecommand \url  [0]{\begingroup\@sanitize@url \@url }%
\providecommand \@url [1]{\endgroup\@href {#1}{\urlprefix }}%
\providecommand \urlprefix  [0]{URL }%
\providecommand \Eprint [0]{\href }%
\providecommand \doibase [0]{http://dx.doi.org/}%
\providecommand \selectlanguage [0]{\@gobble}%
\providecommand \bibinfo  [0]{\@secondoftwo}%
\providecommand \bibfield  [0]{\@secondoftwo}%
\providecommand \translation [1]{[#1]}%
\providecommand \BibitemOpen [0]{}%
\providecommand \bibitemStop [0]{}%
\providecommand \bibitemNoStop [0]{.\EOS\space}%
\providecommand \EOS [0]{\spacefactor3000\relax}%
\providecommand \BibitemShut  [1]{\csname bibitem#1\endcsname}%
\let\auto@bib@innerbib\@empty
\bibitem [{\citenamefont {Ohira}\ and\ \citenamefont
  {Takahara}(2008)}]{Ohira2008}%
  \BibitemOpen
  \bibfield  {author} {\bibinfo {author} {\bibfnamefont {Y.}~\bibnamefont
  {Ohira}}\ and\ \bibinfo {author} {\bibfnamefont {F.}~\bibnamefont
  {Takahara}},\ }\href {\doibase 10.1086/592182} {\bibfield  {journal}
  {\bibinfo  {journal} {The Astrophysical Journal}\ }\textbf {\bibinfo {volume}
  {688}},\ \bibinfo {pages} {320} (\bibinfo {year} {2008})},\ \Eprint
  {http://arxiv.org/abs/0808.3195} {arXiv:0808.3195} \BibitemShut {NoStop}%
\bibitem [{\citenamefont {Buneman}(1963)}]{Buneman1963}%
  \BibitemOpen
  \bibfield  {author} {\bibinfo {author} {\bibfnamefont {O.}~\bibnamefont
  {Buneman}},\ }\href@noop {} {\bibfield  {journal} {\bibinfo  {journal}
  {Physical Review Letters}\ }\textbf {\bibinfo {volume} {10}},\ \bibinfo
  {pages} {285} (\bibinfo {year} {1963})}\BibitemShut {NoStop}%
\bibitem [{\citenamefont {Forslund}\ and\ \citenamefont
  {Shonk}(1970)}]{Forslund1970}%
  \BibitemOpen
  \bibfield  {author} {\bibinfo {author} {\bibfnamefont {D.}~\bibnamefont
  {Forslund}}\ and\ \bibinfo {author} {\bibfnamefont {C.}~\bibnamefont
  {Shonk}},\ }\href {\doibase 10.1103/PhysRevLett.25.281} {\bibfield  {journal}
  {\bibinfo  {journal} {Physical Review Letters}\ }\textbf {\bibinfo {volume}
  {25}},\ \bibinfo {pages} {281} (\bibinfo {year} {1970})}\BibitemShut
  {NoStop}%
\bibitem [{\citenamefont {Karimabadi}\ \emph {et~al.}(1991)\citenamefont
  {Karimabadi}, \citenamefont {Omidi},\ and\ \citenamefont
  {Quest}}]{Karimabadi1991}%
  \BibitemOpen
  \bibfield  {author} {\bibinfo {author} {\bibfnamefont {H.}~\bibnamefont
  {Karimabadi}}, \bibinfo {author} {\bibfnamefont {N.}~\bibnamefont {Omidi}}, \
  and\ \bibinfo {author} {\bibfnamefont {K.~B.}\ \bibnamefont {Quest}},\ }\href
  {\doibase 10.1029/91GL02241} {\bibfield  {journal} {\bibinfo  {journal}
  {Geophysical Research Letters}\ }\textbf {\bibinfo {volume} {18}},\ \bibinfo
  {pages} {1813} (\bibinfo {year} {1991})}\BibitemShut {NoStop}%
\bibitem [{\citenamefont {Akimoto}\ and\ \citenamefont
  {Omidi}(1986)}]{Akimoto1986}%
  \BibitemOpen
  \bibfield  {author} {\bibinfo {author} {\bibfnamefont {K.}~\bibnamefont
  {Akimoto}}\ and\ \bibinfo {author} {\bibfnamefont {N.}~\bibnamefont
  {Omidi}},\ }\href@noop {} {\bibfield  {journal} {\bibinfo  {journal}
  {Geophysical Research Letters}\ }\textbf {\bibinfo {volume} {13}},\ \bibinfo
  {pages} {97} (\bibinfo {year} {1986})}\BibitemShut {NoStop}%
\bibitem [{\citenamefont {Wahlund}\ \emph {et~al.}(1992)\citenamefont
  {Wahlund}, \citenamefont {Forme}, \citenamefont {Opgenoorth}, \citenamefont
  {Persson}, \citenamefont {Mishin},\ and\ \citenamefont
  {Volokitin}}]{Wahlund1992}%
  \BibitemOpen
  \bibfield  {author} {\bibinfo {author} {\bibfnamefont {J.-E.}\ \bibnamefont
  {Wahlund}}, \bibinfo {author} {\bibfnamefont {F.~R.~E.}\ \bibnamefont
  {Forme}}, \bibinfo {author} {\bibfnamefont {H.~J.}\ \bibnamefont
  {Opgenoorth}}, \bibinfo {author} {\bibfnamefont {M.~A.~L.}\ \bibnamefont
  {Persson}}, \bibinfo {author} {\bibfnamefont {E.~V.}\ \bibnamefont {Mishin}},
  \ and\ \bibinfo {author} {\bibfnamefont {A.~S.}\ \bibnamefont {Volokitin}},\
  }\href {\doibase https://doi.org/10.1029/92GL02101} {\bibfield  {journal}
  {\bibinfo  {journal} {Geophysical Research Letters}\ }\textbf {\bibinfo
  {volume} {19}},\ \bibinfo {pages} {1919} (\bibinfo {year}
  {1992})}\BibitemShut {NoStop}%
\bibitem [{\citenamefont {Gr{\'{e}}sillon}\ \emph {et~al.}(1975)\citenamefont
  {Gr{\'{e}}sillon}, \citenamefont {Doveil},\ and\ \citenamefont
  {Buzzi}}]{Gresillon1975}%
  \BibitemOpen
  \bibfield  {author} {\bibinfo {author} {\bibfnamefont {D.}~\bibnamefont
  {Gr{\'{e}}sillon}}, \bibinfo {author} {\bibfnamefont {F.}~\bibnamefont
  {Doveil}}, \ and\ \bibinfo {author} {\bibfnamefont {J.~M.}\ \bibnamefont
  {Buzzi}},\ }\href {\doibase 10.1103/PhysRevLett.34.197} {\bibfield  {journal}
  {\bibinfo  {journal} {Physical Review Letters}\ }\textbf {\bibinfo {volume}
  {34}},\ \bibinfo {pages} {197} (\bibinfo {year} {1975})}\BibitemShut
  {NoStop}%
\bibitem [{\citenamefont {Ohnuma}\ \emph {et~al.}(1976)\citenamefont {Ohnuma},
  \citenamefont {Fujita},\ and\ \citenamefont {Adachi}}]{Ohnuma1976}%
  \BibitemOpen
  \bibfield  {author} {\bibinfo {author} {\bibfnamefont {T.}~\bibnamefont
  {Ohnuma}}, \bibinfo {author} {\bibfnamefont {T.}~\bibnamefont {Fujita}}, \
  and\ \bibinfo {author} {\bibfnamefont {S.}~\bibnamefont {Adachi}},\
  }\href@noop {} {\bibfield  {journal} {\bibinfo  {journal} {Physical Review
  Letters}\ }\textbf {\bibinfo {volume} {36}},\ \bibinfo {pages} {471}
  (\bibinfo {year} {1976})},\ \Eprint {http://arxiv.org/abs/arXiv:1011.1669v3}
  {arXiv:arXiv:1011.1669v3} \BibitemShut {NoStop}%
\bibitem [{\citenamefont {{Takao Fujita}}\ \emph {et~al.}(1977)\citenamefont
  {{Takao Fujita}}, \citenamefont {{Toshiro Ohnuma}},\ and\ \citenamefont
  {{Saburo Adachi}}}]{TakaoFujita1977}%
  \BibitemOpen
  \bibfield  {author} {\bibinfo {author} {\bibnamefont {{Takao Fujita}}},
  \bibinfo {author} {\bibnamefont {{Toshiro Ohnuma}}}, \ and\ \bibinfo {author}
  {\bibnamefont {{Saburo Adachi}}},\ }\href {\doibase
  10.1088/0032-1028/19/9/007} {\bibfield  {journal} {\bibinfo  {journal}
  {Plasma Physics}\ }\textbf {\bibinfo {volume} {19}},\ \bibinfo {pages} {875}
  (\bibinfo {year} {1977})}\BibitemShut {NoStop}%
\bibitem [{\citenamefont {Sarraf}\ \emph {et~al.}(1983)\citenamefont {Sarraf},
  \citenamefont {Williams},\ and\ \citenamefont {Goldman}}]{Sarraf1983}%
  \BibitemOpen
  \bibfield  {author} {\bibinfo {author} {\bibfnamefont {S.~P.}\ \bibnamefont
  {Sarraf}}, \bibinfo {author} {\bibfnamefont {E.~A.}\ \bibnamefont
  {Williams}}, \ and\ \bibinfo {author} {\bibfnamefont {L.~M.}\ \bibnamefont
  {Goldman}},\ }\href@noop {} {\bibfield  {journal} {\bibinfo  {journal}
  {Physical Review A}\ }\textbf {\bibinfo {volume} {27}},\ \bibinfo {pages}
  {2110} (\bibinfo {year} {1983})}\BibitemShut {NoStop}%
\bibitem [{\citenamefont {Ross}\ \emph {et~al.}(2013)\citenamefont {Ross},
  \citenamefont {Park}, \citenamefont {Berger}, \citenamefont {Divol},
  \citenamefont {Kugland}, \citenamefont {Rozmus}, \citenamefont {Ryutov},\
  and\ \citenamefont {Glenzer}}]{Ross2013}%
  \BibitemOpen
  \bibfield  {author} {\bibinfo {author} {\bibfnamefont {J.~S.}\ \bibnamefont
  {Ross}}, \bibinfo {author} {\bibfnamefont {H.-S.}\ \bibnamefont {Park}},
  \bibinfo {author} {\bibfnamefont {R.}~\bibnamefont {Berger}}, \bibinfo
  {author} {\bibfnamefont {L.}~\bibnamefont {Divol}}, \bibinfo {author}
  {\bibfnamefont {N.~L.}\ \bibnamefont {Kugland}}, \bibinfo {author}
  {\bibfnamefont {W.}~\bibnamefont {Rozmus}}, \bibinfo {author} {\bibfnamefont
  {D.}~\bibnamefont {Ryutov}}, \ and\ \bibinfo {author} {\bibfnamefont {S.~H.}\
  \bibnamefont {Glenzer}},\ }\href {\doibase 10.1103/PhysRevLett.110.145005}
  {\bibfield  {journal} {\bibinfo  {journal} {Physical Review Letters}\
  }\textbf {\bibinfo {volume} {110}},\ \bibinfo {pages} {145005} (\bibinfo
  {year} {2013})}\BibitemShut {NoStop}%
\bibitem [{\citenamefont {Rinderknecht}\ \emph {et~al.}(2018)\citenamefont
  {Rinderknecht}, \citenamefont {Park}, \citenamefont {Ross}, \citenamefont
  {Amendt}, \citenamefont {Higginson}, \citenamefont {Wilks}, \citenamefont
  {Haberberger}, \citenamefont {Katz}, \citenamefont {Froula}, \citenamefont
  {Hoffman}, \citenamefont {Kagan}, \citenamefont {Keenan},\ and\ \citenamefont
  {Vold}}]{Rinderknecht2018}%
  \BibitemOpen
  \bibfield  {author} {\bibinfo {author} {\bibfnamefont {H.~G.}\ \bibnamefont
  {Rinderknecht}}, \bibinfo {author} {\bibfnamefont {H.~S.}\ \bibnamefont
  {Park}}, \bibinfo {author} {\bibfnamefont {J.~S.}\ \bibnamefont {Ross}},
  \bibinfo {author} {\bibfnamefont {P.~A.}\ \bibnamefont {Amendt}}, \bibinfo
  {author} {\bibfnamefont {D.~P.}\ \bibnamefont {Higginson}}, \bibinfo {author}
  {\bibfnamefont {S.~C.}\ \bibnamefont {Wilks}}, \bibinfo {author}
  {\bibfnamefont {D.}~\bibnamefont {Haberberger}}, \bibinfo {author}
  {\bibfnamefont {J.}~\bibnamefont {Katz}}, \bibinfo {author} {\bibfnamefont
  {D.~H.}\ \bibnamefont {Froula}}, \bibinfo {author} {\bibfnamefont {N.~M.}\
  \bibnamefont {Hoffman}}, \bibinfo {author} {\bibfnamefont {G.}~\bibnamefont
  {Kagan}}, \bibinfo {author} {\bibfnamefont {B.~D.}\ \bibnamefont {Keenan}}, \
  and\ \bibinfo {author} {\bibfnamefont {E.~L.}\ \bibnamefont {Vold}},\ }\href
  {\doibase 10.1103/PhysRevLett.120.095001} {\bibfield  {journal} {\bibinfo
  {journal} {Physical Review Letters}\ }\textbf {\bibinfo {volume} {120}},\
  \bibinfo {pages} {95001} (\bibinfo {year} {2018})}\BibitemShut {NoStop}%
\bibitem [{\citenamefont {Jiao}\ \emph {et~al.}(2019)\citenamefont {Jiao},
  \citenamefont {He}, \citenamefont {Zhuo}, \citenamefont {Qiao}, \citenamefont
  {Yu}, \citenamefont {Zhang}, \citenamefont {Deng}, \citenamefont {Lu},
  \citenamefont {Zhou}, \citenamefont {Wang}, \citenamefont {Xie},
  \citenamefont {Yang}, \citenamefont {Zhang}, \citenamefont {Zhou},\ and\
  \citenamefont {Gu}}]{Jiao2019}%
  \BibitemOpen
  \bibfield  {author} {\bibinfo {author} {\bibfnamefont {J.~L.}\ \bibnamefont
  {Jiao}}, \bibinfo {author} {\bibfnamefont {S.~K.}\ \bibnamefont {He}},
  \bibinfo {author} {\bibfnamefont {H.~B.}\ \bibnamefont {Zhuo}}, \bibinfo
  {author} {\bibfnamefont {B.}~\bibnamefont {Qiao}}, \bibinfo {author}
  {\bibfnamefont {M.~Y.}\ \bibnamefont {Yu}}, \bibinfo {author} {\bibfnamefont
  {B.}~\bibnamefont {Zhang}}, \bibinfo {author} {\bibfnamefont {Z.~G.}\
  \bibnamefont {Deng}}, \bibinfo {author} {\bibfnamefont {F.}~\bibnamefont
  {Lu}}, \bibinfo {author} {\bibfnamefont {K.~N.}\ \bibnamefont {Zhou}},
  \bibinfo {author} {\bibfnamefont {X.~D.}\ \bibnamefont {Wang}}, \bibinfo
  {author} {\bibfnamefont {N.}~\bibnamefont {Xie}}, \bibinfo {author}
  {\bibfnamefont {L.}~\bibnamefont {Yang}}, \bibinfo {author} {\bibfnamefont
  {F.~Q.}\ \bibnamefont {Zhang}}, \bibinfo {author} {\bibfnamefont {W.~M.}\
  \bibnamefont {Zhou}}, \ and\ \bibinfo {author} {\bibfnamefont {Y.~Q.}\
  \bibnamefont {Gu}},\ }\href {\doibase 10.3847/2041-8213/ab4190} {\bibfield
  {journal} {\bibinfo  {journal} {The Astrophysical Journal}\ }\textbf
  {\bibinfo {volume} {883}},\ \bibinfo {pages} {L37} (\bibinfo {year}
  {2019})}\BibitemShut {NoStop}%
\bibitem [{\citenamefont {Kato}\ and\ \citenamefont
  {Takabe}(2010)}]{Kato2010b}%
  \BibitemOpen
  \bibfield  {author} {\bibinfo {author} {\bibfnamefont {T.~N.}\ \bibnamefont
  {Kato}}\ and\ \bibinfo {author} {\bibfnamefont {H.}~\bibnamefont {Takabe}},\
  }\href {\doibase 10.1063/1.3372138} {\bibfield  {journal} {\bibinfo
  {journal} {Physics of Plasmas}\ }\textbf {\bibinfo {volume} {17}},\ \bibinfo
  {pages} {032114} (\bibinfo {year} {2010})}\BibitemShut {NoStop}%
\bibitem [{\citenamefont {Sarri}\ \emph {et~al.}(2011)\citenamefont {Sarri},
  \citenamefont {Dieckmann}, \citenamefont {Kourakis},\ and\ \citenamefont
  {Borghesi}}]{Sarri2011b}%
  \BibitemOpen
  \bibfield  {author} {\bibinfo {author} {\bibfnamefont {G.}~\bibnamefont
  {Sarri}}, \bibinfo {author} {\bibfnamefont {M.~E.}\ \bibnamefont
  {Dieckmann}}, \bibinfo {author} {\bibfnamefont {I.}~\bibnamefont {Kourakis}},
  \ and\ \bibinfo {author} {\bibfnamefont {M.}~\bibnamefont {Borghesi}},\
  }\href {\doibase 10.1103/PhysRevLett.107.025003} {\bibfield  {journal}
  {\bibinfo  {journal} {Physical Review Letters}\ }\textbf {\bibinfo {volume}
  {107}},\ \bibinfo {pages} {025003} (\bibinfo {year} {2011})}\BibitemShut
  {NoStop}%
\bibitem [{\citenamefont {Zhang}\ \emph {et~al.}(2018)\citenamefont {Zhang},
  \citenamefont {Cai},\ and\ \citenamefont {Zhu}}]{Zhang2018}%
  \BibitemOpen
  \bibfield  {author} {\bibinfo {author} {\bibfnamefont {W.-s.}\ \bibnamefont
  {Zhang}}, \bibinfo {author} {\bibfnamefont {H.-b.}\ \bibnamefont {Cai}}, \
  and\ \bibinfo {author} {\bibfnamefont {S.-p.}\ \bibnamefont {Zhu}},\
  }\href@noop {} {\bibfield  {journal} {\bibinfo  {journal} {Plasma Physics and
  Controlled Fusion}\ }\textbf {\bibinfo {volume} {60}},\ \bibinfo {pages}
  {055001} (\bibinfo {year} {2018})}\BibitemShut {NoStop}%
\bibitem [{\citenamefont {Denavit}(1992)}]{Denavit1992}%
  \BibitemOpen
  \bibfield  {author} {\bibinfo {author} {\bibfnamefont {J.}~\bibnamefont
  {Denavit}},\ }\href {\doibase 10.1103/PhysRevLett.69.3052} {\bibfield
  {journal} {\bibinfo  {journal} {Physical Review Letters}\ }\textbf {\bibinfo
  {volume} {69}},\ \bibinfo {pages} {3052} (\bibinfo {year}
  {1992})}\BibitemShut {NoStop}%
\bibitem [{\citenamefont {Silva}\ \emph {et~al.}(2004)\citenamefont {Silva},
  \citenamefont {Marti}, \citenamefont {Davies}, \citenamefont {Fonseca},
  \citenamefont {Ren}, \citenamefont {Tsung},\ and\ \citenamefont
  {Mori}}]{Silva2004}%
  \BibitemOpen
  \bibfield  {author} {\bibinfo {author} {\bibfnamefont {L.~O.}\ \bibnamefont
  {Silva}}, \bibinfo {author} {\bibfnamefont {M.}~\bibnamefont {Marti}},
  \bibinfo {author} {\bibfnamefont {J.~R.}\ \bibnamefont {Davies}}, \bibinfo
  {author} {\bibfnamefont {R.~A.}\ \bibnamefont {Fonseca}}, \bibinfo {author}
  {\bibfnamefont {C.}~\bibnamefont {Ren}}, \bibinfo {author} {\bibfnamefont
  {F.}~\bibnamefont {Tsung}}, \ and\ \bibinfo {author} {\bibfnamefont {W.~B.}\
  \bibnamefont {Mori}},\ }\href {\doibase 10.1103/PhysRevLett.92.015002}
  {\bibfield  {journal} {\bibinfo  {journal} {Physical Review Letters}\
  }\textbf {\bibinfo {volume} {92}},\ \bibinfo {pages} {015002} (\bibinfo
  {year} {2004})}\BibitemShut {NoStop}%
\bibitem [{\citenamefont {Fiuza}\ \emph {et~al.}(2012)\citenamefont {Fiuza},
  \citenamefont {Stockem}, \citenamefont {Boella}, \citenamefont {Fonseca},
  \citenamefont {Silva}, \citenamefont {Haberberger}, \citenamefont
  {Tochitsky}, \citenamefont {Gong}, \citenamefont {Mori},\ and\ \citenamefont
  {Joshi}}]{Fiuza2012}%
  \BibitemOpen
  \bibfield  {author} {\bibinfo {author} {\bibfnamefont {F.}~\bibnamefont
  {Fiuza}}, \bibinfo {author} {\bibfnamefont {A.}~\bibnamefont {Stockem}},
  \bibinfo {author} {\bibfnamefont {E.}~\bibnamefont {Boella}}, \bibinfo
  {author} {\bibfnamefont {R.~A.}\ \bibnamefont {Fonseca}}, \bibinfo {author}
  {\bibfnamefont {L.~O.}\ \bibnamefont {Silva}}, \bibinfo {author}
  {\bibfnamefont {D.}~\bibnamefont {Haberberger}}, \bibinfo {author}
  {\bibfnamefont {S.}~\bibnamefont {Tochitsky}}, \bibinfo {author}
  {\bibfnamefont {C.}~\bibnamefont {Gong}}, \bibinfo {author} {\bibfnamefont
  {W.~B.}\ \bibnamefont {Mori}}, \ and\ \bibinfo {author} {\bibfnamefont
  {C.}~\bibnamefont {Joshi}},\ }\href {\doibase 10.1103/PhysRevLett.109.215001}
  {\bibfield  {journal} {\bibinfo  {journal} {Physical Review Letters}\
  }\textbf {\bibinfo {volume} {109}},\ \bibinfo {pages} {215001} (\bibinfo
  {year} {2012})}\BibitemShut {NoStop}%
\bibitem [{\citenamefont {Kumar}\ \emph {et~al.}(2019)\citenamefont {Kumar},
  \citenamefont {Sakawa}, \citenamefont {D{\"{o}}hl}, \citenamefont {Woolsey},\
  and\ \citenamefont {Morace}}]{Kumar2019a}%
  \BibitemOpen
  \bibfield  {author} {\bibinfo {author} {\bibfnamefont {R.}~\bibnamefont
  {Kumar}}, \bibinfo {author} {\bibfnamefont {Y.}~\bibnamefont {Sakawa}},
  \bibinfo {author} {\bibfnamefont {L.~N.}\ \bibnamefont {D{\"{o}}hl}},
  \bibinfo {author} {\bibfnamefont {N.}~\bibnamefont {Woolsey}}, \ and\
  \bibinfo {author} {\bibfnamefont {A.}~\bibnamefont {Morace}},\ }\href
  {\doibase 10.1103/PhysRevAccelBeams.22.043401} {\bibfield  {journal}
  {\bibinfo  {journal} {Physical Review Accelerators and Beams}\ }\textbf
  {\bibinfo {volume} {22}},\ \bibinfo {pages} {043401} (\bibinfo {year}
  {2019})}\BibitemShut {NoStop}%
\bibitem [{\citenamefont {Haberberger}\ \emph {et~al.}(2011)\citenamefont
  {Haberberger}, \citenamefont {Tochitsky}, \citenamefont {Fiuza},
  \citenamefont {Gong}, \citenamefont {Fonseca}, \citenamefont {Silva},
  \citenamefont {Mori},\ and\ \citenamefont {Joshi}}]{Haberberger2011}%
  \BibitemOpen
  \bibfield  {author} {\bibinfo {author} {\bibfnamefont {D.}~\bibnamefont
  {Haberberger}}, \bibinfo {author} {\bibfnamefont {S.}~\bibnamefont
  {Tochitsky}}, \bibinfo {author} {\bibfnamefont {F.}~\bibnamefont {Fiuza}},
  \bibinfo {author} {\bibfnamefont {C.}~\bibnamefont {Gong}}, \bibinfo {author}
  {\bibfnamefont {R.~A.}\ \bibnamefont {Fonseca}}, \bibinfo {author}
  {\bibfnamefont {L.~O.}\ \bibnamefont {Silva}}, \bibinfo {author}
  {\bibfnamefont {W.~B.}\ \bibnamefont {Mori}}, \ and\ \bibinfo {author}
  {\bibfnamefont {C.}~\bibnamefont {Joshi}},\ }\href {\doibase
  10.1038/nphys2130} {\bibfield  {journal} {\bibinfo  {journal} {Nature
  Physics}\ }\textbf {\bibinfo {volume} {8}},\ \bibinfo {pages} {95} (\bibinfo
  {year} {2011})}\BibitemShut {NoStop}%
\bibitem [{\citenamefont {Tresca}\ \emph {et~al.}(2015)\citenamefont {Tresca},
  \citenamefont {Dover}, \citenamefont {Cook}, \citenamefont {Maharjan},
  \citenamefont {Polyanskiy}, \citenamefont {Najmudin}, \citenamefont
  {Shkolnikov},\ and\ \citenamefont {Pogorelsky}}]{Tresca2015}%
  \BibitemOpen
  \bibfield  {author} {\bibinfo {author} {\bibfnamefont {O.}~\bibnamefont
  {Tresca}}, \bibinfo {author} {\bibfnamefont {N.~P.}\ \bibnamefont {Dover}},
  \bibinfo {author} {\bibfnamefont {N.}~\bibnamefont {Cook}}, \bibinfo {author}
  {\bibfnamefont {C.}~\bibnamefont {Maharjan}}, \bibinfo {author}
  {\bibfnamefont {M.~N.}\ \bibnamefont {Polyanskiy}}, \bibinfo {author}
  {\bibfnamefont {Z.}~\bibnamefont {Najmudin}}, \bibinfo {author}
  {\bibfnamefont {P.}~\bibnamefont {Shkolnikov}}, \ and\ \bibinfo {author}
  {\bibfnamefont {I.}~\bibnamefont {Pogorelsky}},\ }\href {\doibase
  10.1103/PhysRevLett.115.094802} {\bibfield  {journal} {\bibinfo  {journal}
  {Physical Review Letters}\ }\textbf {\bibinfo {volume} {115}},\ \bibinfo
  {pages} {094802} (\bibinfo {year} {2015})}\BibitemShut {NoStop}%
\bibitem [{\citenamefont {Zhang}\ \emph {et~al.}(2015)\citenamefont {Zhang},
  \citenamefont {Shen}, \citenamefont {Wang}, \citenamefont {Xu}, \citenamefont
  {Liu}, \citenamefont {Liang}, \citenamefont {Leng}, \citenamefont {Li},
  \citenamefont {Yan}, \citenamefont {Chen},\ and\ \citenamefont
  {Xu}}]{Zhang2015}%
  \BibitemOpen
  \bibfield  {author} {\bibinfo {author} {\bibfnamefont {H.}~\bibnamefont
  {Zhang}}, \bibinfo {author} {\bibfnamefont {B.~F.}\ \bibnamefont {Shen}},
  \bibinfo {author} {\bibfnamefont {W.~P.}\ \bibnamefont {Wang}}, \bibinfo
  {author} {\bibfnamefont {Y.}~\bibnamefont {Xu}}, \bibinfo {author}
  {\bibfnamefont {Y.~Q.}\ \bibnamefont {Liu}}, \bibinfo {author} {\bibfnamefont
  {X.~Y.}\ \bibnamefont {Liang}}, \bibinfo {author} {\bibfnamefont {Y.~X.}\
  \bibnamefont {Leng}}, \bibinfo {author} {\bibfnamefont {R.~X.}\ \bibnamefont
  {Li}}, \bibinfo {author} {\bibfnamefont {X.~Q.}\ \bibnamefont {Yan}},
  \bibinfo {author} {\bibfnamefont {J.~E.}\ \bibnamefont {Chen}}, \ and\
  \bibinfo {author} {\bibfnamefont {Z.~Z.}\ \bibnamefont {Xu}},\ }\href
  {\doibase 10.1063/1.4907194} {\bibfield  {journal} {\bibinfo  {journal}
  {Physics of Plasmas}\ }\textbf {\bibinfo {volume} {22}},\ \bibinfo {pages}
  {013113} (\bibinfo {year} {2015})}\BibitemShut {NoStop}%
\bibitem [{\citenamefont {Zhang}\ \emph {et~al.}(2017)\citenamefont {Zhang},
  \citenamefont {Shen}, \citenamefont {Wang}, \citenamefont {Zhai},
  \citenamefont {Li}, \citenamefont {Lu}, \citenamefont {Li}, \citenamefont
  {Xu}, \citenamefont {Wang}, \citenamefont {Liang}, \citenamefont {Leng},
  \citenamefont {Li},\ and\ \citenamefont {Xu}}]{Zhang2017}%
  \BibitemOpen
  \bibfield  {author} {\bibinfo {author} {\bibfnamefont {H.}~\bibnamefont
  {Zhang}}, \bibinfo {author} {\bibfnamefont {B.~F.}\ \bibnamefont {Shen}},
  \bibinfo {author} {\bibfnamefont {W.~P.}\ \bibnamefont {Wang}}, \bibinfo
  {author} {\bibfnamefont {S.~H.}\ \bibnamefont {Zhai}}, \bibinfo {author}
  {\bibfnamefont {S.~S.}\ \bibnamefont {Li}}, \bibinfo {author} {\bibfnamefont
  {X.~M.}\ \bibnamefont {Lu}}, \bibinfo {author} {\bibfnamefont {J.~F.}\
  \bibnamefont {Li}}, \bibinfo {author} {\bibfnamefont {R.~J.}\ \bibnamefont
  {Xu}}, \bibinfo {author} {\bibfnamefont {X.~L.}\ \bibnamefont {Wang}},
  \bibinfo {author} {\bibfnamefont {X.~Y.}\ \bibnamefont {Liang}}, \bibinfo
  {author} {\bibfnamefont {Y.~X.}\ \bibnamefont {Leng}}, \bibinfo {author}
  {\bibfnamefont {R.~X.}\ \bibnamefont {Li}}, \ and\ \bibinfo {author}
  {\bibfnamefont {Z.~Z.}\ \bibnamefont {Xu}},\ }\href {\doibase
  10.1103/PhysRevLett.119.164801} {\bibfield  {journal} {\bibinfo  {journal}
  {Physical Review Letters}\ }\textbf {\bibinfo {volume} {119}},\ \bibinfo
  {pages} {164801} (\bibinfo {year} {2017})}\BibitemShut {NoStop}%
\bibitem [{\citenamefont {Antici}\ \emph {et~al.}(2017)\citenamefont {Antici},
  \citenamefont {Boella}, \citenamefont {Chen}, \citenamefont {Andrews},
  \citenamefont {Barberio}, \citenamefont {B{\"{o}}ker}, \citenamefont
  {Cardelli}, \citenamefont {Feugeas}, \citenamefont {Glesser}, \citenamefont
  {Nicola{\"{i}}}, \citenamefont {Romagnani}, \citenamefont {Scisci{\`{o}}},
  \citenamefont {Starodubtsev}, \citenamefont {Willi}, \citenamefont {Kieffer},
  \citenamefont {Tikhonchuk}, \citenamefont {P{\'{e}}pin}, \citenamefont
  {Silva}, \citenamefont {Humi{\`{e}}res},\ and\ \citenamefont
  {Fuchs}}]{Antici2017}%
  \BibitemOpen
  \bibfield  {author} {\bibinfo {author} {\bibfnamefont {P.}~\bibnamefont
  {Antici}}, \bibinfo {author} {\bibfnamefont {E.}~\bibnamefont {Boella}},
  \bibinfo {author} {\bibfnamefont {S.~N.}\ \bibnamefont {Chen}}, \bibinfo
  {author} {\bibfnamefont {D.~S.}\ \bibnamefont {Andrews}}, \bibinfo {author}
  {\bibfnamefont {M.}~\bibnamefont {Barberio}}, \bibinfo {author}
  {\bibfnamefont {J.}~\bibnamefont {B{\"{o}}ker}}, \bibinfo {author}
  {\bibfnamefont {F.}~\bibnamefont {Cardelli}}, \bibinfo {author}
  {\bibfnamefont {J.~L.}\ \bibnamefont {Feugeas}}, \bibinfo {author}
  {\bibfnamefont {M.}~\bibnamefont {Glesser}}, \bibinfo {author} {\bibfnamefont
  {P.}~\bibnamefont {Nicola{\"{i}}}}, \bibinfo {author} {\bibfnamefont
  {L.}~\bibnamefont {Romagnani}}, \bibinfo {author} {\bibfnamefont
  {M.}~\bibnamefont {Scisci{\`{o}}}}, \bibinfo {author} {\bibfnamefont
  {M.}~\bibnamefont {Starodubtsev}}, \bibinfo {author} {\bibfnamefont
  {O.}~\bibnamefont {Willi}}, \bibinfo {author} {\bibfnamefont {J.~C.}\
  \bibnamefont {Kieffer}}, \bibinfo {author} {\bibfnamefont {V.}~\bibnamefont
  {Tikhonchuk}}, \bibinfo {author} {\bibfnamefont {H.}~\bibnamefont
  {P{\'{e}}pin}}, \bibinfo {author} {\bibfnamefont {L.~O.}\ \bibnamefont
  {Silva}}, \bibinfo {author} {\bibfnamefont {E.~D.}\ \bibnamefont
  {Humi{\`{e}}res}}, \ and\ \bibinfo {author} {\bibfnamefont {J.}~\bibnamefont
  {Fuchs}},\ }\href {\doibase 10.1038/s41598-017-15449-8} {\bibfield  {journal}
  {\bibinfo  {journal} {Scientific Reports}\ }\textbf {\bibinfo {volume} {7}},\
  \bibinfo {pages} {1} (\bibinfo {year} {2017})},\ \Eprint
  {http://arxiv.org/abs/1708.02539} {arXiv:1708.02539} \BibitemShut {NoStop}%
\bibitem [{\citenamefont {Pak}\ \emph {et~al.}(2018)\citenamefont {Pak},
  \citenamefont {Kerr}, \citenamefont {Lemos}, \citenamefont {Link},
  \citenamefont {Patel}, \citenamefont {Albert}, \citenamefont {Divol},
  \citenamefont {Pollock}, \citenamefont {Haberberger}, \citenamefont {Froula},
  \citenamefont {Gauthier}, \citenamefont {Glenzer}, \citenamefont {Longman},
  \citenamefont {Manzoor}, \citenamefont {Fedosejevs}, \citenamefont
  {Tochitsky}, \citenamefont {Joshi},\ and\ \citenamefont {Fiuza}}]{Pak2018a}%
  \BibitemOpen
  \bibfield  {author} {\bibinfo {author} {\bibfnamefont {A.}~\bibnamefont
  {Pak}}, \bibinfo {author} {\bibfnamefont {S.}~\bibnamefont {Kerr}}, \bibinfo
  {author} {\bibfnamefont {N.}~\bibnamefont {Lemos}}, \bibinfo {author}
  {\bibfnamefont {A.}~\bibnamefont {Link}}, \bibinfo {author} {\bibfnamefont
  {P.}~\bibnamefont {Patel}}, \bibinfo {author} {\bibfnamefont
  {F.}~\bibnamefont {Albert}}, \bibinfo {author} {\bibfnamefont
  {L.}~\bibnamefont {Divol}}, \bibinfo {author} {\bibfnamefont {B.~B.}\
  \bibnamefont {Pollock}}, \bibinfo {author} {\bibfnamefont {D.}~\bibnamefont
  {Haberberger}}, \bibinfo {author} {\bibfnamefont {D.}~\bibnamefont {Froula}},
  \bibinfo {author} {\bibfnamefont {M.}~\bibnamefont {Gauthier}}, \bibinfo
  {author} {\bibfnamefont {S.~H.}\ \bibnamefont {Glenzer}}, \bibinfo {author}
  {\bibfnamefont {A.}~\bibnamefont {Longman}}, \bibinfo {author} {\bibfnamefont
  {L.}~\bibnamefont {Manzoor}}, \bibinfo {author} {\bibfnamefont
  {R.}~\bibnamefont {Fedosejevs}}, \bibinfo {author} {\bibfnamefont
  {S.}~\bibnamefont {Tochitsky}}, \bibinfo {author} {\bibfnamefont
  {C.}~\bibnamefont {Joshi}}, \ and\ \bibinfo {author} {\bibfnamefont
  {F.}~\bibnamefont {Fiuza}},\ }\href {\doibase
  10.1103/PhysRevAccelBeams.21.103401} {\bibfield  {journal} {\bibinfo
  {journal} {Physical Review Accelerators and Beams}\ }\textbf {\bibinfo
  {volume} {21}},\ \bibinfo {pages} {103401} (\bibinfo {year} {2018})},\
  \Eprint {http://arxiv.org/abs/1810.08190} {arXiv:1810.08190} \BibitemShut
  {NoStop}%
\bibitem [{\citenamefont {Ota}\ \emph {et~al.}(2019)\citenamefont {Ota},
  \citenamefont {Morace}, \citenamefont {Kumar}, \citenamefont {Kambayashi},
  \citenamefont {Egashira}, \citenamefont {Kanasaki}, \citenamefont {Fukuda},\
  and\ \citenamefont {Sakawa}}]{Ota2019}%
  \BibitemOpen
  \bibfield  {author} {\bibinfo {author} {\bibfnamefont {M.}~\bibnamefont
  {Ota}}, \bibinfo {author} {\bibfnamefont {A.}~\bibnamefont {Morace}},
  \bibinfo {author} {\bibfnamefont {R.}~\bibnamefont {Kumar}}, \bibinfo
  {author} {\bibfnamefont {S.}~\bibnamefont {Kambayashi}}, \bibinfo {author}
  {\bibfnamefont {S.}~\bibnamefont {Egashira}}, \bibinfo {author}
  {\bibfnamefont {M.}~\bibnamefont {Kanasaki}}, \bibinfo {author}
  {\bibfnamefont {Y.}~\bibnamefont {Fukuda}}, \ and\ \bibinfo {author}
  {\bibfnamefont {Y.}~\bibnamefont {Sakawa}},\ }\href {\doibase
  10.1016/j.hedp.2019.100697} {\bibfield  {journal} {\bibinfo  {journal} {High
  Energy Density Physics}\ }\textbf {\bibinfo {volume} {33}},\ \bibinfo {pages}
  {100697} (\bibinfo {year} {2019})}\BibitemShut {NoStop}%
\bibitem [{\citenamefont {Bulanov}\ \emph {et~al.}(2014)\citenamefont
  {Bulanov}, \citenamefont {Wilkens}, \citenamefont {Esirkepov}, \citenamefont
  {Korn}, \citenamefont {Kraft}, \citenamefont {Kraft}, \citenamefont {Molls},\
  and\ \citenamefont {Khoroshkov}}]{Bulanov2014a}%
  \BibitemOpen
  \bibfield  {author} {\bibinfo {author} {\bibfnamefont {S.~V.}\ \bibnamefont
  {Bulanov}}, \bibinfo {author} {\bibfnamefont {J.~J.}\ \bibnamefont
  {Wilkens}}, \bibinfo {author} {\bibfnamefont {T.~Z.}\ \bibnamefont
  {Esirkepov}}, \bibinfo {author} {\bibfnamefont {G.}~\bibnamefont {Korn}},
  \bibinfo {author} {\bibfnamefont {G.}~\bibnamefont {Kraft}}, \bibinfo
  {author} {\bibfnamefont {S.~D.}\ \bibnamefont {Kraft}}, \bibinfo {author}
  {\bibfnamefont {M.}~\bibnamefont {Molls}}, \ and\ \bibinfo {author}
  {\bibfnamefont {V.}~\bibnamefont {Khoroshkov}},\ }\href
  {http://ufn.ru/en/articles/2014/12/a/references.html} {\bibfield  {journal}
  {\bibinfo  {journal} {Physics-Uspekhi}\ }\textbf {\bibinfo {volume} {57}},\
  \bibinfo {pages} {1149} (\bibinfo {year} {2014})}\BibitemShut {NoStop}%
\bibitem [{\citenamefont {Grismayer}\ and\ \citenamefont
  {Mora}(2006)}]{Grismayer2006}%
  \BibitemOpen
  \bibfield  {author} {\bibinfo {author} {\bibfnamefont {T.}~\bibnamefont
  {Grismayer}}\ and\ \bibinfo {author} {\bibfnamefont {P.}~\bibnamefont
  {Mora}},\ }\href {\doibase 10.1063/1.2178653} {\bibfield  {journal} {\bibinfo
   {journal} {Physics of Plasmas}\ }\textbf {\bibinfo {volume} {13}},\ \bibinfo
  {pages} {032103} (\bibinfo {year} {2006})}\BibitemShut {NoStop}%
\bibitem [{\citenamefont {Kumar}\ \emph {et~al.}(2021)\citenamefont {Kumar},
  \citenamefont {Sakawa}, \citenamefont {Sano}, \citenamefont {Dohl},
  \citenamefont {Woolsey},\ and\ \citenamefont {Morace}}]{Kumar2021}%
  \BibitemOpen
  \bibfield  {author} {\bibinfo {author} {\bibfnamefont {R.}~\bibnamefont
  {Kumar}}, \bibinfo {author} {\bibfnamefont {Y.}~\bibnamefont {Sakawa}},
  \bibinfo {author} {\bibfnamefont {T.}~\bibnamefont {Sano}}, \bibinfo {author}
  {\bibfnamefont {L.~N.~K.}\ \bibnamefont {Dohl}}, \bibinfo {author}
  {\bibfnamefont {N.}~\bibnamefont {Woolsey}}, \ and\ \bibinfo {author}
  {\bibfnamefont {A.}~\bibnamefont {Morace}},\ }\href {\doibase
  10.1103/PhysRevE.103.043201} {\bibfield  {journal} {\bibinfo  {journal}
  {Physical Review E}\ }\textbf {\bibinfo {volume} {103}},\ \bibinfo {pages}
  {43201} (\bibinfo {year} {2021})},\ \Eprint {http://arxiv.org/abs/2104.00866}
  {arXiv:2104.00866} \BibitemShut {NoStop}%
\bibitem [{\citenamefont {Arber}\ \emph {et~al.}(2015)\citenamefont {Arber},
  \citenamefont {Bennett}, \citenamefont {Brady}, \citenamefont
  {Lawrence-Douglas}, \citenamefont {Ramsay}, \citenamefont {Sircombe},
  \citenamefont {Gillies}, \citenamefont {Evans}, \citenamefont {Schmitz},
  \citenamefont {Bell},\ and\ \citenamefont {Ridgers}}]{Arber2015}%
  \BibitemOpen
  \bibfield  {author} {\bibinfo {author} {\bibfnamefont {T.~D.}\ \bibnamefont
  {Arber}}, \bibinfo {author} {\bibfnamefont {K.}~\bibnamefont {Bennett}},
  \bibinfo {author} {\bibfnamefont {C.~S.}\ \bibnamefont {Brady}}, \bibinfo
  {author} {\bibfnamefont {A.}~\bibnamefont {Lawrence-Douglas}}, \bibinfo
  {author} {\bibfnamefont {M.~G.}\ \bibnamefont {Ramsay}}, \bibinfo {author}
  {\bibfnamefont {N.~J.}\ \bibnamefont {Sircombe}}, \bibinfo {author}
  {\bibfnamefont {P.}~\bibnamefont {Gillies}}, \bibinfo {author} {\bibfnamefont
  {R.~G.}\ \bibnamefont {Evans}}, \bibinfo {author} {\bibfnamefont
  {H.}~\bibnamefont {Schmitz}}, \bibinfo {author} {\bibfnamefont {A.~R.}\
  \bibnamefont {Bell}}, \ and\ \bibinfo {author} {\bibfnamefont {C.~P.}\
  \bibnamefont {Ridgers}},\ }\href {\doibase 10.1088/0741-3335/57/11/113001}
  {\bibfield  {journal} {\bibinfo  {journal} {Plasma Physics and Controlled
  Fusion}\ }\textbf {\bibinfo {volume} {57}} (\bibinfo {year} {2015}),\
  10.1088/0741-3335/57/11/113001}\BibitemShut {NoStop}%
\bibitem [{\citenamefont {Tidman}\ and\ \citenamefont {Krall}(1971)}]{Tidman}%
  \BibitemOpen
  \bibfield  {author} {\bibinfo {author} {\bibfnamefont {D.~A.}\ \bibnamefont
  {Tidman}}\ and\ \bibinfo {author} {\bibfnamefont {N.~A.}\ \bibnamefont
  {Krall}},\ }\href@noop {} {\emph {\bibinfo {title} {{Shock Waves in
  Collisionless Plasmas}}}}\ (\bibinfo  {publisher} {Wiley-Interscience},\
  \bibinfo {address} {New York},\ \bibinfo {year} {1971})\BibitemShut {NoStop}%
\bibitem [{\citenamefont {Fried}\ and\ \citenamefont
  {Conte}(1961)}]{Fried1961book}%
  \BibitemOpen
  \bibfield  {author} {\bibinfo {author} {\bibfnamefont {B.~D.}\ \bibnamefont
  {Fried}}\ and\ \bibinfo {author} {\bibfnamefont {S.~D.}\ \bibnamefont
  {Conte}},\ }\href@noop {} {\emph {\bibinfo {title} {{The Plasma Dispersion
  Function}}}}\ (\bibinfo  {publisher} {Academic Press},\ \bibinfo {address}
  {New York},\ \bibinfo {year} {1961})\BibitemShut {NoStop}%
\bibitem [{\citenamefont {Ichimaru}(1992)}]{Ichimaru1992}%
  \BibitemOpen
  \bibfield  {author} {\bibinfo {author} {\bibfnamefont {S.}~\bibnamefont
  {Ichimaru}},\ }\href@noop {} {\emph {\bibinfo {title} {{Statistical Plasma
  Physics, Volume I: Basic Principles}}}}\ (\bibinfo  {publisher}
  {Addison-Wesley},\ \bibinfo {year} {1992})\BibitemShut {NoStop}%
\bibitem [{\citenamefont {Kruer}\ and\ \citenamefont
  {Estabrook}(1985)}]{Kruer1985}%
  \BibitemOpen
  \bibfield  {author} {\bibinfo {author} {\bibfnamefont {W.~L.}\ \bibnamefont
  {Kruer}}\ and\ \bibinfo {author} {\bibfnamefont {K.}~\bibnamefont
  {Estabrook}},\ }\href {\doibase 10.1063/1.865171} {\bibfield  {journal}
  {\bibinfo  {journal} {Physics of Fluids}\ }\textbf {\bibinfo {volume} {28}},\
  \bibinfo {pages} {430} (\bibinfo {year} {1985})}\BibitemShut {NoStop}%
\bibitem [{\citenamefont {Mora}(2003)}]{Mora2003}%
  \BibitemOpen
  \bibfield  {author} {\bibinfo {author} {\bibfnamefont {P.}~\bibnamefont
  {Mora}},\ }\href {\doibase 10.1103/PhysRevLett.90.185002} {\bibfield
  {journal} {\bibinfo  {journal} {Physical Review Letters}\ }\textbf {\bibinfo
  {volume} {90}},\ \bibinfo {pages} {185002} (\bibinfo {year}
  {2003})}\BibitemShut {NoStop}%
\end{thebibliography}
%

\end{document}